\documentclass[aip,jcp,superscriptaddress,reprint]{revtex4-1}

\usepackage[utf8]{inputenc}
\usepackage[T1]{fontenc}
\usepackage{graphicx}
\usepackage{bm}
\usepackage{amsmath}
\usepackage{amssymb}
\usepackage{mathtools}
\usepackage[version=4]{mhchem}
\usepackage{siunitx}
\usepackage{microtype}
\usepackage{float}
\usepackage{booktabs}

\usepackage[colorlinks, allcolors=blue]{hyperref}
\hypersetup{pdftitle={Force Training by Taylor Expansion},
            pdfauthor={A.~Cooper, J.~Kaestner, A.~Urban, N.~Artrith}}

\makeatletter
\let\Hy@backout\@gobble
\makeatother

\let\onlinecite\citenum

\begin{document}

\title{Efficient Training of ANN Potentials by Including Atomic Forces
  via Taylor Expansion and Application to Water and a Transition-Metal
  Oxide}

\author{April M.~Cooper}%
\affiliation{School of Chemistry, University of St Andrews, UK}%
\affiliation{Institute for Theoretical Chemistry, University of Stuttgart, Germany}%
\author{Johannes Kästner}%
\affiliation{Institute for Theoretical Chemistry, University of Stuttgart, Germany}%
\author{Alexander Urban}%
\affiliation{School of Chemistry, University of St Andrews, UK}%
\affiliation{\hbox{Department of Chemical Engineering, Columbia University, New York, NY, USA}}%
\author{Nongnuch Artrith}%
\email{nartrith@atomistic.net}%
\affiliation{School of Chemistry, University of St Andrews, UK}%
\affiliation{\hbox{Department of Chemical Engineering, Columbia University, New York, NY, USA}}%
\date{\today}

\begin{abstract}
  Artificial neural network (ANN) potentials enable the efficient
  large-scale atomistic modeling of complex materials with near
  first-principles accuracy.
  For molecular dynamics simulations, accurate energies and interatomic
  forces are a prerequisite, but training ANN potentials simultaneously
  on energies and forces from electronic structure calculations is
  computationally demanding.
  Here, we introduce an efficient alternative method for the training of
  ANN potentials on energy and force information based on an
  extrapolation of the total energy via a Taylor expansion.
  By translating the force information to approximate energies, the
  quadratic scaling with the number of atoms exhibited by conventional
  force-training methods can be avoided, which enables the training on
  reference data sets containing complex atomic structures.
  We demonstrate for different materials systems, clusters of water
  molecules, bulk liquid water, and a lithium transition-metal oxide,
  that the proposed force-training approach provides substantial
  improvements over schemes that train on energies only.
  Including force information for training reduces the size of the
  reference data sets required for ANN potential construction, increases
  the transferability of the potential, and generally improves the force
  prediction accuracy.
  For a set of water clusters, the Taylor-expansion approach achieves
  around 50\% of the force error improvement compared to the explicit
  training on all force components, at a much smaller computational
  cost.
  The alternative force training approach thus simplifies the
  construction of general ANN potentials for the prediction of accurate
  energies and interatomic forces for diverse types of materials, as
  demonstrated here for water and a transition-metal oxide.

\end{abstract}

\maketitle


\section{Introduction}
\label{sec:introduction}

During the past decade, machine-learning potentials (MLP) trained on
accurate first principles and quantum chemistry methods have become an
integral part of computational materials
science.\cite{prl98-2007-146401, prl104-2010-136403, jocp-2014-SNAP,
  ijqc115-2015-1051, acie56-2017-12828, ncm4-2018-48, as-2019-1900808,
  ncm5-2019-51, ncm5-2019-55, ncm5-2019-75, ncm5-2019-83}
Carefully constructed MLPs can be as accurate as their first principles
reference method but at a fraction of the computational cost and with an
effort that scales linearly with the number of atoms, which enables the
modeling of complex materials that are not accessible with first
principles methods such as density-functional theory
(DFT).\cite{pr136-64-846, pr140-65-A1133, jcp136-2012-150901,
  rmp87-2015-897}
MLPs can also be trained on results from highly accurate ab-initio
methods such as coupled-cluster calculations.\cite{jpca123-2019-9061,
  nc10-2019-2903, jctc16-2019-88}

One popular MLP variant is the high dimensional artificial neural
network (ANN) potential method initially introduced by Behler and
Parrinello for elemental Si in 2007\cite{prl98-2007-146401} and extended
to multiple chemical species by Artrith and
coworkers,\cite{prB83-2011-153101, prb96-2017-014112} exploring also the
effect of long-ranged electrostatic
interactions.\cite{prB83-2011-153101}
ANN potentials have been successfully used to model many complex
materials, such as elemental metals,\cite{prB85-2012-045439,
  jotacs140-2018-2812} alloys,\cite{prb85-2012-174103, nl14-2014-2670}
oxides,\cite{acsc6-2016-1675} molecular systems,\cite{pnas113-2016-8368,
  cs8-2017-3192, jcp148-2018-094106, tjopcl10-2019-6067} amorphous
phases,\cite{tjocp147-2017-214106, jcp148-2018-241711, cm30-2019-7077,
  arxiv1901-2019-09272} interfaces,\cite{pssb250-2013-1191,
  pccp18-2016-28704, tjocp148-2018-241720, jpe1-2019-ML-for-Interfaces}
and nanoporous materials.\cite{joctac15-2019-3793}

Once trained, the computational cost of ANN potentials does not scale
with the number of data points used for training, so that training sets
can be as large as necessary to sample the relevant and potentially
diverse chemical and structural space.
Often, ANN potentials are trained on total energy information, i.e., a
single piece of information per DFT calculation is used as reference for
the potential training.
As a result, a large number of DFT reference calculations, possibly tens
to hundreds of thousands,\cite{arxiv1901-2019-09272} may be needed to
achieve the desired interpolation accuracy for applications in Monte
Carlo (MC) sampling or molecular dynamics (MD) simulations.

\begin{figure*}
  \linespread{1.0}
  \centering
  \includegraphics[width=0.8\textwidth]{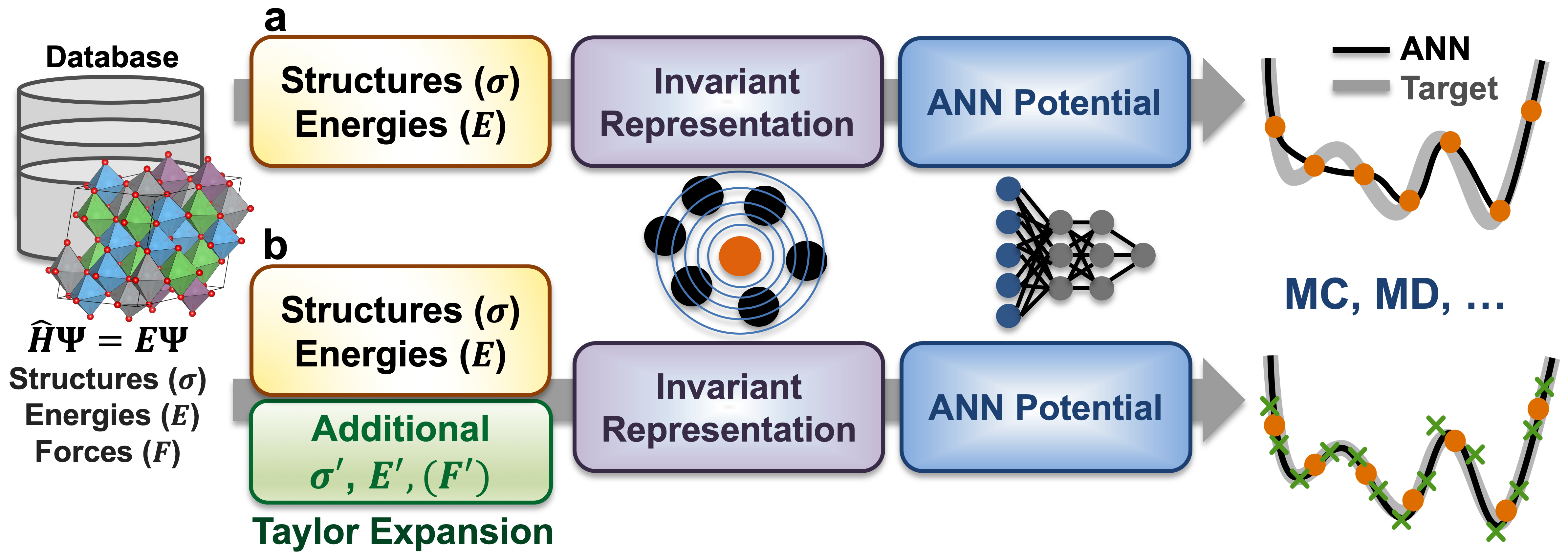}
  \caption{\label{fig:schematic}%
    \textbf{Flowchart of the construction of artificial neural network
      (ANN) potentials.} \textbf{a,} Often, ANN potentials are trained
    on total energies from reference electronic structure calculations,
    since direct training of gradients (interatomic forces) is
    computationally demanding.  This approach requires large numbers of
    reference calculations to converge the slope of the potential energy
    surface. \textbf{b,} In this work, a new scheme for ANN potential
    training is introduced, in which interatomic forces are used to
    extend the training set with energies approximated by Taylor
    expansion.  The method is computationally as efficient as the
    conventional energy training and can significantly reduce the number
    of required reference calculations.}
\end{figure*}
Large training sets are needed to accurately capture the gradient of the
potential energy surface (PES) and thus the interatomic forces.
An accurate representation of the atomic forces is crucial for MD
simulations and geometry optimizations, and hence the force prediction
error is an important target parameter for potential construction.
At the same time, the energy is what determines the most stable
structure or phase and other materials properties, and \emph{energy
  conservation} needs to be obeyed by any interatomic potential, so that
an accurate representation of the structural energy is also needed.

In ANN potential training, the force prediction accuracy is typically
converged by increasing the number of reference structures used for
training until the relevant structure space is sampled sufficiently fine
to also represent the gradient of the PES.\cite{prB85-2012-045439,
  pssb250-2013-1191}
This strategy does not only increase the training set size it also makes
the training technically more challenging since relevant structures have
to be carefully selected without adding redundancies to the training
set.

In principle, atomic forces or higher derivatives of the energy from
first principles can also be used as reference data for ANN potential
training.
To train on force information, the \emph{loss function} for the ANN
potential training has to include the force prediction error, i.e., the
error of the negative gradient of the energy.
However, including the gradient in the loss function introduces a
significant computational overhead because training requires the
gradient of the loss function and thus the second derivative of the ANN
potential is needed (i.e., the \emph{Hessian} matrix).
Therefore, in practice sometimes hybrid approaches are used in which the
atomic forces of a subset of atoms are included in the loss
function.\cite{jctc15-2019-3075}
Such approaches are especially useful in combination with online
training methods that allow the selection of different force components
for each training iteration (epoch).

In the present article, we introduce a new scheme for including atomic
force information in the ANN potential training that avoids the
computationally demanding evaluation of higher order derivatives (see
flowchart in \textbf{Figure~\ref{fig:schematic}}).
The approach, which is based on a Taylor extrapolation of the total
energy, is detailed in the following \emph{Background} section.
In the \emph{Results} section, we first demonstrate the basic principle
of the methodology for an analytical example and then apply it to
systems with increasing complexity (clusters of water molecules, bulk
water, and a quaternary metal oxide), finding that a smaller number of
reference structures compared to energy-only training is sufficient for
converging the force error in ANN potential construction.

\section{Background}
\label{sec:background}

\vspace{-\baselineskip}

\subsection{Artificial neural network potentials}
\label{sec:ANN-pot}

\vspace{-\baselineskip}

Artificial neural network (ANN) potentials are a type of many-body
interatomic potential for atomistic simulations.\cite{prl98-2007-146401,
  acie56-2017-12828, cms114-2016-135}
In contrast to conventional potentials that are based on an approximate
representation of the physical atomic interactions, such as embedded
atom models,\cite{prB29-1984-6443} ANN potentials employ general
flexible functions, ANNs, for the interpolation between reference data
points from first-principles calculations.
ANN potentials represent the total energy $E(\sigma)$ of an atomic
structure $\sigma=\{\vec{R}_{i}\}$ ($\vec{R}_{i}$ is the position
vector of atom $i$) as the sum of atomic energies

\vspace{-\baselineskip}
\begin{align}
  E(\sigma)
  \approx E^{\textsc{ann}}(\sigma)
  = \sum_{i}E_{\textup{atom}}(\sigma^{R_{c}}_{i})
  \label{eq:atomic-energy}
\end{align}
where $\sigma^{R_{c}}_{i}$ is the local atomic environment of atom $i$,
i.e., the atomic positions and chemical species of all atoms within a
given cutoff radius $R_{c}$ of atom $i$.\cite{prl98-2007-146401,
  acie56-2017-12828, cms114-2016-135}
The atomic energy function $E_{\textup{atom}}$ in
equation~\eqref{eq:atomic-energy} is given by an ANN specific for each
chemical species $t$ (i.e., $t$ is the \emph{type} of atom $i$)

\vspace{-\baselineskip}
\begin{align}
  E_{\textup{atom}}(\sigma^{R_{c}}_{i})
  = \textup{ANN}_{t}(\widetilde{\sigma}^{R_{c}}_{i})
  \quad\text{for atom type } t
  \quad .
\end{align}

ANNs require an input of constant size, but the number of atoms within
the local atomic environment of an atom $i$, $\sigma^{R_{c}}_{i}$, can
vary.
A suitable input feature vector is obtained by transforming
$\sigma^{R_{c}}_{i}$ to a descriptor $\widetilde{\sigma}^{R_{c}}_{i}$
with constant dimension that is also invariant with respect to
(i)~translation/rotation of the entire structure and (ii)~exchange of
equivalent atoms.
In the present work, we employed the Chebyshev descriptors by Artrith,
Urban, and Ceder,\cite{prb96-2017-014112} and the symmetry function
descriptor by Behler and Parrinello.\cite{prl98-2007-146401,
  jcp134-2011-074106}

Further details of the ANN architecture and the descriptor parameters
are given in the methods section~\ref{sec:methods}.

\subsection{ANN potential training with reference energies}
\label{sec:energy-training}

ANN potentials are trained to reproduce the structural energies of
reference data sets containing atomic structures $\sigma$ (input) and
energies $E(\sigma)$ (output) from first principles calculations.
The atomic energy $E_{\textup{atom}}$ is not uniquely defined from first
principles, so that no reference for the direct training of
$\textup{ANN}_{t}(\widetilde{\sigma}^{R_{c}}_{i})$ is available.
Note that there are non-unique approaches that can be used to decompose
the total structural energy from first principles into atomic
contributions.\cite{prb99-2019-064103}
To avoid ambiguities and additional model complexity, it is more
straightforward to implement the ANN potential training such that it
minimizes a loss function based on the total energy $E(\sigma)$

\vspace{-\baselineskip}
\begin{align}
\begin{aligned}
  \mathcal{L}
  &= \sum_{\sigma} \frac{1}{2} \Bigl[\Delta E(\sigma)\Bigr]^{2}
  \\
  \text{with}\;\;
  \Delta E(\sigma) &= E^{\textsc{ann}}(\sigma) - E(\sigma)
 \label{eq:loss-function-no-F}
  \quad ,
\end{aligned}
\end{align}
where $E^{\textsc{ann}}(\sigma)$ is the energy of structure $\sigma$
predicted by the ANN potential of equation~\eqref{eq:atomic-energy}.

Gradient-based optimization methods require the derivative of the loss
function with respect to the ANN parameters, the ANN \emph{weights}
$\{w_{k}\}$

\vspace{-\baselineskip}
\begin{align}
\begin{aligned}
  \frac{\partial}{\partial w_{k}} \mathcal{L}
  &= \sum_{\sigma} \Delta E(\sigma)
     \frac{\partial}{\partial w_{k}} E^{\textsc{ann}}(\sigma)
  \\
  &= \sum_{\sigma} \Delta E(\sigma)
  \sum_{i\in\sigma} \frac{\partial}{\partial w_{k}} \textup{ANN}_{t}(\widetilde{\sigma}^{R_{c}}_{i})
  \quad ,
\end{aligned}
\end{align}
where the weight derivatives of the ANNs can be obtained using the
standard backpropagation method.\cite{OrrMueller1998_BackProp}

The computational complexity of backpropagation scales as
$\mathcal{O}(N_{w})$ where $N_{w}$ is the number of weight parameters.
For a training set containing a total of $N_{\textup{atom}}$ atoms, the
computational cost of one training epoch is therefore proportional to
$\mathcal{O}(N_{\textup{atom}}N_{w})$.

\subsection{Training with reference atomic forces}
\label{sec:force-training}

Density-functional theory (DFT)\cite{pr136-64-846, pr140-65-A1133}
calculations with local or semi-local density functionals provide the
interatomic forces at minimal overhead.
Considering the importance of accurate forces for structure
optimizations and MD simulations, it is desirable to include the force
error in the loss function $\mathcal{L}$ of
equation~\eqref{eq:loss-function-no-F}.
In addition, each atomic structure has only one total energy but
3~atomic force components \emph{per atom}.
Hence, using the atomic forces as additional reference data increases
the data points available for training per atomic structure by a factor
of $3\,N$, where $N$ is the number of atoms in the structure.

The Cartesian vector of the force acting on atom $i$ is given by the
negative gradient of the total energy

\vspace{-\baselineskip}
\begin{align}
  \vec{F}_{i}(\sigma)
  = - \vec{\nabla}_{i} E(\sigma)
  \quad\text{with}\quad
  \vec{\nabla}_{i}
  = \frac{\partial}{\partial\vec{R}_{i}}
  \label{eq:atomic-force}
\end{align}
where $\vec{R}_{i}=(R^{x}_{i}, R^{y}_{i}, R^{z}_{i})^{T}$ is the
Cartesian position vector of atom $i$.
Including the atomic forces in training can thus be accomplished with
the loss function

\vspace{-\baselineskip}
\begin{align}
\begin{aligned}
  \mathcal{L}'
  &= \mathcal{L}
  + a\sum_{\sigma} \frac{1}{2} \Bigl[
    \sum_{j} \Bigl(
      -\vec{\nabla}_{j} E^{\textsc{ann}}(\sigma) - \vec{F}_{j}(\sigma)
    \Bigr)
  \Bigr]^{2} \\
  &= \mathcal{L}
  + a\sum_{\sigma} \frac{1}{2} \Bigl[
    \Delta \vec{F}(\sigma)
  \Bigr]^{2}
  \\
  \text{with}&\quad
  \Delta \vec{F}(\sigma)
  = \sum_{j} \Bigl(-\vec{\nabla}_{j} E^{\textsc{ann}}(\sigma) - \vec{F}_{j}(\sigma)\Bigr)
  \label{eq:error-function-F}
  \quad ,
\end{aligned}
\end{align}
where $a$ is an adjustable parameter that determines the contribution of
the force error to the loss function $\mathcal{L}$.\cite{Lorenz2001}
The gradient of the new loss function $\mathcal{L}'$ with respect to
the weight parameters is

\vspace{-\baselineskip}
\begin{widetext}
\begin{align}
  \frac{\partial}{\partial w_{k}} \mathcal{L}'
  &= \frac{\partial}{\partial w_{k}} \mathcal{L}
  - a\sum_{\sigma} \Bigl[
    \sum_{j\in\sigma} \Bigl( -\vec{\nabla}_{i} E^{\textsc{ann}}(\sigma) - \vec{F}_{j} \Bigr)
  \Bigr]
  \sum_{j\in\sigma}\frac{\partial}{\partial w_{k}} \vec{\nabla}_{j} E^{\textsc{ann}}(\sigma)
  \label{eq:weight-gradient-F}
  \quad .
\end{align}
\end{widetext}
As seen in equation~\eqref{eq:weight-gradient-F}, evaluating the weight
gradient of the new loss function $\mathcal{L}'$ requires taking the
derivative of the position gradient of the ANN potential

\vspace{-\baselineskip}
\begin{align}
  \sum_{j\in\sigma}\frac{\partial}{\partial w_{k}} \vec{\nabla}_{j}  E^{\textsc{ann}}(\sigma)
  = \sum_{j\in\sigma}\sum_{i\in\sigma}
  \frac{\partial}{\partial w_{k}} \vec{\nabla}_{j}
  \textup{ANN}_{t}(\widetilde{\sigma}^{R_{c}}_{i})
\end{align}
which scales quadratic with the number of atoms in the reference
structure.
Note that only the cross terms for atoms within two times the cutoff
radius of the potential ($2R_{c}$) are different from
zero.\cite{prB85-2012-045439}
For very large or periodic structures the scaling of the derivative
evaluation therefore eventually becomes linear but with a large
pre-factor of $N_{\textup{local}}$ that is the average number of atoms
within $2R_{c}$ and can be several hundred to thousands depending on the
density of the material and the cutoff radius.

The quadratic scaling and the often large number of atoms in reference
data sets typically makes it \emph{infeasible} to train all force
components using the loss function in
equation~\eqref{eq:error-function-F}.
One option is to include a fraction of all force components in the loss
function, e.g., only 0.41\% of the atomic forces were included in
reference \onlinecite{jctc15-2019-3075}, and Artrith et al.\ have
previously used only 10\% or less of the force information for ANN
potential training.\cite{prB83-2011-153101, prB85-2012-045439,
  pssb250-2013-1191}
This strategy can work especially well with online training methods that
allow selecting different force components at each epoch which is not
possible using batch training methods.

The importance of atomic force information and the unfavorable scaling
of direct force training prompted us to consider alternative means of
including force information in ANN potential training.

\subsection{Including atomic force information via Taylor expansion}
\label{sec:taylor-expansion}

Training of the energy with the loss function of
equation~\eqref{eq:loss-function-no-F} is efficient, so that a
translation of the force information to (approximate) energy information
is advantageous.
Such a translation can be accomplished using a first-order Taylor
expansion to estimate the energy of additionally generated atomic
structures without the need to perform additional electronic structure
calculations.

The energy of a structure $\sigma'=\{\vec{R}'_{i}\}$ that was generated by
displacing the atoms in the original structure $\sigma=\{\vec{R}_{i}\}$
can be expressed as the Taylor series

\vspace{-\baselineskip}
\begin{align}
\begin{aligned}
  &E(\sigma')
  = E(\sigma)
  + \sum_{i} \vec{\delta}_{i} \vec{\nabla}_{i}E(\sigma)
  \\
  &\quad+ \frac{1}{2}\sum_{i} \vec{\delta}_{i}^{2} \vec{\nabla}^{2}_{i}E(\sigma)
  + \ldots
  \;\text{with}\;
  \vec{\delta}_{i} = \vec{R}'_{i} - \vec{R}_{i}
  \label{eq:Taylor-series}
  \quad .
\end{aligned}
\end{align}
Substituting the atomic force of equation~\eqref{eq:atomic-force} for
the negative gradient of the energy and truncating after the first
order, we arrive at the approximation

\vspace{-\baselineskip}
\begin{align}
  E(\sigma')
  \approx E(\sigma) - \sum_{i} \vec{\delta}_{i} \vec{F}_{i}(\sigma)
  \label{eq:1st-order-expansion}
  \quad ,
\end{align}
where $E(\sigma)$ and $\vec{F}_{i}(\sigma)$ are the energies and atomic
forces from the original reference electronic structure calculations.
The first-order approximation is valid for small displacements
$\{\vec{\delta}_{i}\}$.

Equation~\eqref{eq:1st-order-expansion} provides a mechanism for
incorporating approximate force information in the ANN potential
training as additional structure-energy pair records.
What remains is to decide on a recipe for generating such additional
structures by atom displacement.
In the present work, we considered two strategies:
(A)~displacing one individual atom to generate one additional structure,
and (B)~randomly displacing all atoms.
In strategy (A), we select an atom $i$ in structure $\sigma$ and
displace its coordinates by a small amount $\pm\delta$ in each Cartesian
direction to generate 6~additional structures.
For example, displacing atom $i$ in structure
$\sigma={\vec{R}_{1}, \vec{R}_{2}, \ldots, \vec{R}_{N}}$ in the negative
Cartesian $x$ direction would yield the new structure--energy pair
\begin{equation}
  \begin{aligned}
  \sigma' &= \{\vec{R}_{1}, \ldots, \vec{R}_{i}-\delta\hat{x}, \ldots, \vec{R}_{N}\}\\
  E(\sigma') &= E(\sigma) - \delta F^{x}_{i}(\sigma)
  \end{aligned}
  \quad ,
\end{equation}
where $\hat{x}$ is the unit vector in the Cartesian $x$ direction, and
$F_{i}^{x}$ is the $x$ component of the force acting on atom $i$.
A similar approach has previously been used by Vlcek et al.\ for
molecular force-field optimization with energy and force
information.\cite{jcp147-2017-161713}

In strategy (B), all atoms are displaced by small random vectors
$\vec{\delta}_{i}$ where the total displacement is such that
$|\vec{\delta}_{i}|\leq{}\delta_{\textup{max}}$ for all atoms.
Since a net translation of the entire structure does not affect the
energy, the center-of-mass displacement is subtracted from the combined
atomic displacements.
The energy of the resulting structure is evaluated according to
equation~\eqref{eq:1st-order-expansion}.

The optimal values of the displacement $\delta$ in strategy (A) and the
maximal displacement $\delta_{\textup{max}}$ in strategy (B) are
parameters and will be determined in the following.


\section{Results}
\label{results}

To assess the efficacy of the Taylor-expansion approach laid out in the
previous section, we considered a series of materials systems with
increasing complexity:
An analytic Lennard-Jones dimer molecule as test case, clusters of water
molecules, a periodic bulk water box, and a complex oxide system with
five different chemical species.

\subsection{Diatomic Molecule}
\label{sec:dimer}

\begin{figure*}
  \linespread{1.0}
  \centering
  \includegraphics[width=0.8\textwidth]{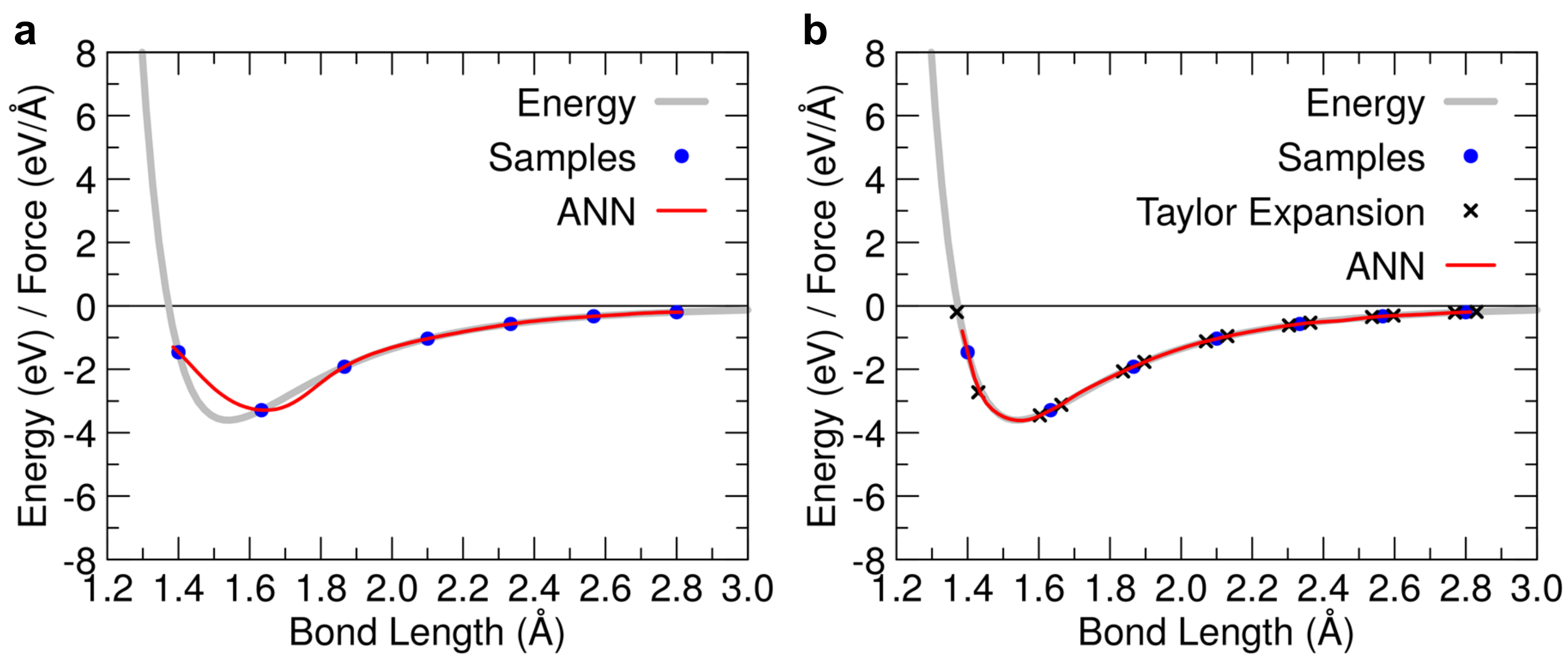}
  \caption{\label{fig:dimer}%
    \textbf{Analytic dimer example illustrating the Taylor expansion
      approach.}
    The gray lines indicate the energy vs.~bond length curve of the
    model Lennard-Jones potential discussed in the text, and the
    7~samples used for ANN potential training are shown as blue
    circles. The red line in panel \textbf{a}~corresponds to an ANN
    potential trained only on the energy of the reference data points
    (blue circles).  In panel \textbf{b}~the ANN potential (red line)
    was additionally trained on 14~displaced structures (black crosses)
    with displacements of $\delta=\pm{}0.02$~\AA{}, the energies of
    which were approximated by a first-order Taylor expansion using the
    analytic gradient at the blue reference data points.}
\end{figure*}
Our first test case is a diatomic molecule with atomic interactions
described by the analytic Lennard-Jones potential\cite{tfs25-1929-668}

\vspace{-\baselineskip}
\begin{align}
  V(r) = \varepsilon\,\Biggl[
         \Bigl(\frac{r_{0}}{r}\Bigr)^{12}
      - 2\Bigl(\frac{r_{0}}{r}\Bigr)^{6}
  \Biggr]
\end{align}
with binding energy $\varepsilon=3.607$~eV and equilibrium distance
$r_{0}=1.54$~\AA{}.
The binding energy and bond distance were chosen such that they
correspond to a typical covalent bond, approximately the carbon-carbon
bond, so that the magnitude of the displacement in the Taylor expansion
will be comparable to an actual compound.
An analytic interatomic potential has the advantage that the gradient
and the Taylor expansion can be calculated analytically.
Further, the one-dimensional dimer molecule makes it straightforward to
visualize the PES, i.e., the bond-energy curve, so that the analytic
potential and the ANN interpolation can be visually compared with each
other.

\textbf{Figure~\ref{fig:dimer}a} shows the bond energy in a range close
to the equilibrium distance and an ANN potential that was trained on
seven equidistant reference points between 1.4~\AA{} and 2.8~\AA{}.
As seen in the figure, the ANN potential approximates the Lennard-Jones
potential well at distances above 1.8~\AA{}, but is unable to reproduce
the minimum region and predicts an incorrect larger equilibrium distance
as well as an incorrect slope and curvature near the minimum.
The result of training the ANN potential with the Taylor expansion
approach is shown in \textbf{Figure~\ref{fig:dimer}b}, where additional
data points were generated by approximating the energy for slightly
longer and slightly shorter bond distances using the first-order Taylor
expansion of equation~\eqref{eq:1st-order-expansion} using a
displacement of $\delta=$~0.02~\AA{}.
As seen in the figure, the additional approximate data points guide the
ANN potential, and the resulting fit is visually much better and
reproduces the correct bond minimum as well as the slope and the
curvature within the minimum well.

Also apparent in \textbf{Figure~\ref{fig:dimer}b} are the deviations of
the first-order Taylor expansion from the true bond energy curve.
The approach is approximate and will give rise to noise in the reference
data, and the optimal amount of additional approximate data points per
exact energy data point has to be determined such that the accuracy of
the ANN potential does not suffer.
In the following, we will assess the Taylor expansion method for real
materials systems to quantify both the improvement of the force
prediction accuracy and the effect of noise in the reference energies.

\subsection{Water clusters}
\label{sec:water-clusters}

While the analytic dimer example is useful for illustrative purposes,
our objective is the training on first-principles energies and forces.
As a second test case, we therefore consider a reference data set based
on structures obtained from MD simulations of water clusters with six
water molecules.
MD simulations at 300~K and 800~K were performed on the semiempirical
\emph{Geometry, Frequency, Noncovalent, eXtended TB}
(GFN-xTB)\cite{jctc13-2017-1989} level of theory.
The atomic forces and energies of a subset of the structures along the
MD trajectories were recalculated using a first-principles
density-functional theory (DFT) approach (BLYP-D3/def2-TZVP) to be used
as reference data for ANN potential construction.
See the methods section~\ref{sec:DFT} for the details of our MD and DFT
calculations.

As a baseline for the assessment of the force-training method, we first
quantify the error in the atomic forces if only total energies are
trained using the methodology and loss function of
section~\ref{sec:energy-training}.
As a second point of reference the error in the atomic forces is
quantified for ANN potentials that were trained by including the error
in the atomic forces in the loss function as described in
section~\ref{sec:force-training}.

\begin{figure*}
  \linespread{1.0}
  \centering
  \includegraphics[width=\textwidth]{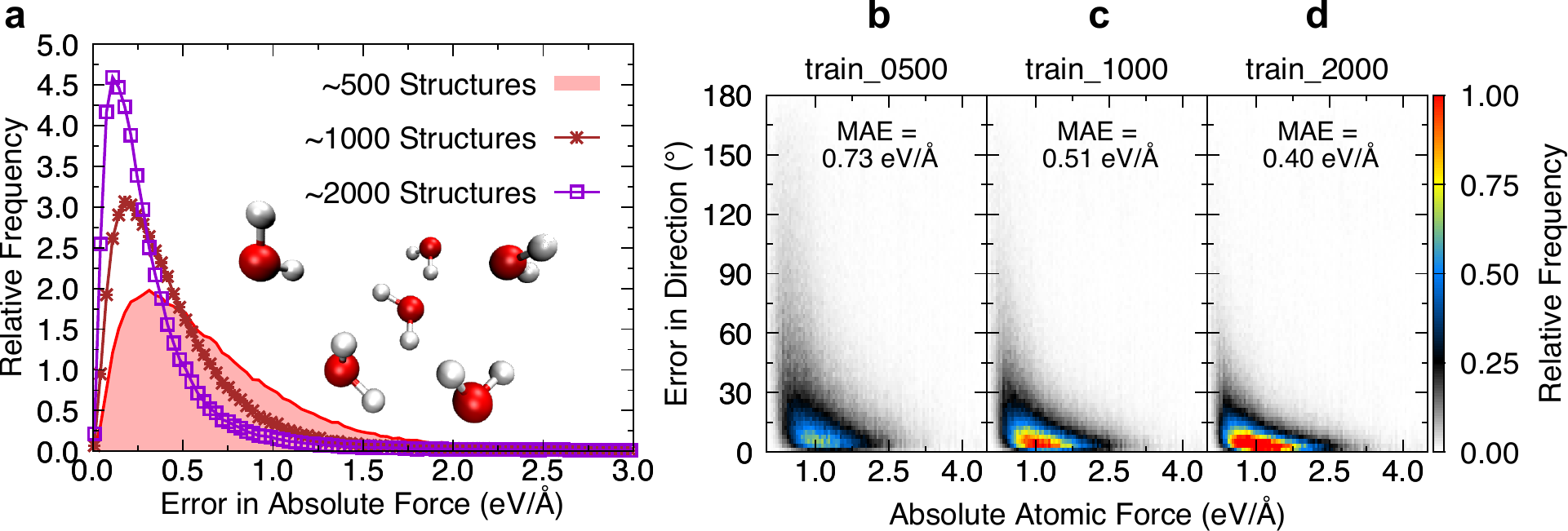}
  \caption{\label{fig:trnset-size}%
    \textbf{Impact of the training set size on the ANN force prediction
      errors for clusters of six water molecules.}  \textbf{a},~Force
    error distribution after training on the three different training
    sets with $\sim$500 (\texttt{train\_0500}, red shaded), $\sim$1000
    (\texttt{train\_1000}, dark red stars), and $\sim$2000
    (\texttt{train\_2000}, violet squares) structures.  The inset shows
    a representative water cluster structure.  \textbf{b-d},~Frequency
    of occurrence of a given error in the direction of the force in
    degrees as a function of the absolute atomic force for energy
    training and varying training set size.  Panels \textbf{b-d} also
    shows the mean absolute errors (MEA) of the predicted forces for the
    three different training sets.  All statistics shown are based on
    10~ANN potentials trained for each scenario.  }
\end{figure*}

\subsubsection{Total energy training with increasingly large training
  sets}

As mentioned in the introduction, a common way to improve the force
prediction error of ANN potentials is by increasing the reference data
set size to sample the configurational space more finely.
We therefore investigated first the influence of the size of the
reference data set on the quality of the force prediction of ANN
potentials that were trained on the total energy only.

In order to study how the size of the reference data set affects the
quality of the force prediction, three different reference data sets
with increasing number of data points were assembled from the MD
reference data set:
\begin{itemize}
\item[(\textit{i})] a subset containing 471~reference structures
  referred to as the \texttt{train\_0500} data set in the following,
\item[(\textit{ii})] a set with 943 structures (\texttt{train\_1000}),
  and
\item[(\textit{iii})] a set with 1,886~structures (\texttt{train\_2000}).
\end{itemize}
The structures within these subsets were chosen evenly spaced along the
MD trajectories to ensure a maximum decorrelation of the reference data.

For each training set, the ANN potential training was repeated ten times
with different random initial weight parameters $\{w_{k}\}$ to obtain
statistics on the prediction quality of the atomic forces for the
resulting ANN potentials.
Details on the ANN potential training are given in section
\ref{sec:ANN-parameters}.
All errors reported in the following were obtained for the same
independent validation data set containing 2000 structures that are
not included in any of the training sets.

\textbf{Figure~\ref{fig:trnset-size}a} shows the distributions of the
error in the norm of the predicted atomic forces for the structures
within the validation set after training on the energies in the
\texttt{train\_0500}, \texttt{train\_1000}, and \texttt{train\_2000} data
sets.
Only the \emph{energy} error entered the loss function, so that the
interatomic \emph{forces} were not directly trained.

As seen in the figure, for the smallest \texttt{train\_0500} set
interpolating the ANN PES using only the energies of the reference
structures leads to a wide distribution of errors in the prediction of
the absolute value of the atomic forces.
Especially, the pronounced tail of the distribution implies that there is
a large fraction of atoms for which the absolute value of the force is
predicted with an error greater than 1.0~eV/\AA{}.
Increasing the size of the training set to $\sim$1000 and $\sim$2000
structures reduces the tail of the error distribution significantly.

\textbf{Figures~\ref{fig:trnset-size}b-d} show a corresponding analysis
of the error in the direction of the predicted atomic forces and the
mean absolute error (MAE) of the atomic forces.
High relative frequencies of occurrence are shown in yellow and red,
whereas low relative frequencies are colored in shades of gray.
The errors are shown as a function of the absolute value of the atomic
force, and it can be seen that the reliability of the prediction of the
direction of the atomic forces increases for increasing absolute values
of the force vectors.
This means, the error in the force direction is greater for small force
vectors than for large force vectors.
Especially for atoms with atomic forces with absolute values of less
than 1.0~eV/\AA{}, the direction of the force vector predicted by the ANN
potential scatters strongly.

This scattering is significantly decreased for atomic forces with large
absolute values.
As seen in panel~\textbf{a}, for the small \texttt{train\_0500} set the
error distribution has a shallow maximum between 0~and~25$^{\circ}$
depending on the absolute force value, but the heat map shows much
larger errors of nearly 180$^{\circ}$ for force vectors with small
absolute value.

Additionally, increasing the size of the reference data set reduces the
scattering in the predictions notably.
Particularly, the scattering in the prediction of atomic forces with
absolute values smaller than 1.0~eV is notably decreased in comparison
to the results obtained from the \texttt{train\_0500} reference data
set.
Furthermore, the number of atoms for which the error is
$\SI{15}{\degree}$ or less increases significantly when the ANN
potential is trained with the \texttt{train\_2000} reference data
instead of the \texttt{train\_0500} data set.

Depending on the reference method and the size of the atomic structures,
increasing the number of structures in the training set may entail a
massive computational overhead.
Therefore, we next investigate whether the Taylor-expansion formalism
of section~\ref{sec:taylor-expansion} could alleviate the need for large
training set sizes.

\subsubsection{Optimal meta-parameters for the Taylor-expansion approach}

Next, we investigate the efficacy of approximate force training using
the Taylor-expansion approach of section~\ref{sec:taylor-expansion}.
In order to apply the two displacement strategies, i.e., (A)~the
displacement of single atoms in the three Cartesian directions and
(B)~the displacement of all atoms in random directions, suitable
displacement parameters $\delta$ and $\delta_{\textup{max}}$ have to be
determined.
Finally, the optimal number of additional structures with approximate
energies to be generated using the Taylor-expansion approach needs to be
determined, as the computational effort for ANN potential training
scales with the size of the training set.

The number of additional structures is given in terms of a multiple $a$
of the original training set, so that $a=X$ means that the number of
generated structures is $X$ times the number of the original structures.
For example, the \texttt{train\_0500} data set contains 471 structures,
so that for $a=10$ a total of $10\times{}471=4,710$ additional
structures with approximate energy will be generated by atomic
displacement.

For each possible choice of a parameter pair $(a,\delta)$ or
$(a,\delta_{\textup{max}})$ ten ANN potentials were trained to obtain
statistics on the resulting ANN potentials.
The optimal parameter pair was chosen by calculating the MAE of the
atomic forces for the validation data set and thereby averaging the
errors obtained from all ten potentials fitted for each parameter pair.

\begin{figure}
  \linespread{1.0}
  \centering
  \includegraphics[width=\columnwidth]{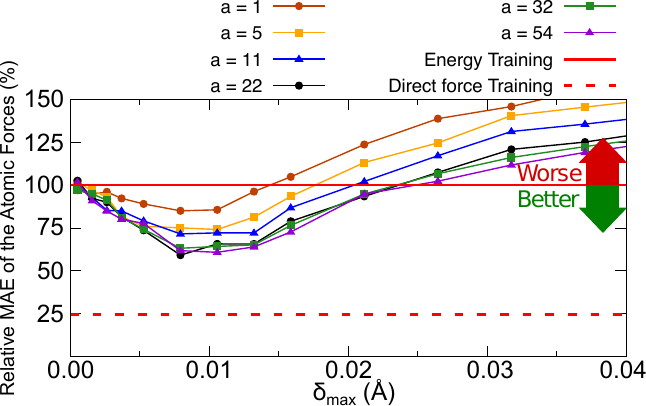}
  \caption{\label{fig:displacement-and-fraction}%
    \textbf{Relative mean absolute error (MAE) of the atomic forces for
      different force-training parameters.}  The MAE is shown as a
    function of the atomic displacement $\delta_\textup{max}$ for
    different multiples $a$ of additionally generated structures.  The
    graph shows the MAE relative to the DFT reference.  Results are
    shown for displacement strategy (B) and for the smallest reference
    data set (\texttt{train\_0500}).  The red dashed line indicates the
    reduction of the force error that can be achieved by direct force
    training.}
\end{figure}
\textbf{Figure~\ref{fig:displacement-and-fraction}} shows the MAE of the
atomic forces obtained after training with the Taylor-expansion approach
as function of the maximum displacement $\delta_\textup{max}$ for
different multiples of additional structures $a$.
The ANN potentials used to quantify this error were fitted to the
\texttt{train\_0500} reference data set using random displacement
strategy~(B) for the given $\delta_\textup{max}$ parameters.
In the figure, the MAE are given relative to the one obtained if no
force information is used for training.

As seen in \textbf{Figure~\ref{fig:displacement-and-fraction}}, the
values of $\delta_\textup{max}$ that lead to the greatest reduction of
the MAE are independent of the number of additionally generated
structures $a$ and are close to $\delta_\textup{max}=0.01$~\AA{}.
If the displacement parameter is chosen too large, the first-order
Taylor expansion is no longer a good approximation and the force error
increases.
Additionally, the MAE decreases with increasing number of generated
structures until a multiple of $a=22$ has been reached.
For multiples $a$ greater than 22 no significant further improvement was
found.
Based on this analysis we conclude that, for the \texttt{train\_0500}
reference data set, the optimal parameter choice that leads to the
smallest MAE with the Taylor-expansion approach is $a=22$ and
$\delta_{\textup{max}}= \SI{0.008}{\angstrom}$.

The optimization of the meta-parameters was repeated for the larger
training sets \texttt{train\_1000} and \texttt{train\_2000} and for the
second Taylor-expansion displacement strategy (A), and the optimal
parameters are given in \textbf{Table~S1} in the supplementary
information.
In summary, the optimal displacement $\delta$ and multiple $a$ do not
appear to depend on the size of the reference data set.
Out of the considered values, the optimal multiple of generated
structures is $a=22$, and optimal displacements are
$\delta_{(A)} = 0.03$~\AA{} for strategy~(A) and
$\delta_{(B)} = 0.008$~\AA{} for strategy~(B).
The heat maps in supplementary \textbf{Figure~S1} show the similarity of
the error in the force direction for varying displacements.

\subsubsection{Energy and force training with the Taylor-expansion approach}

\begin{figure*}
  \linespread{1.0}
  \centering
  \includegraphics[width=0.8\textwidth]{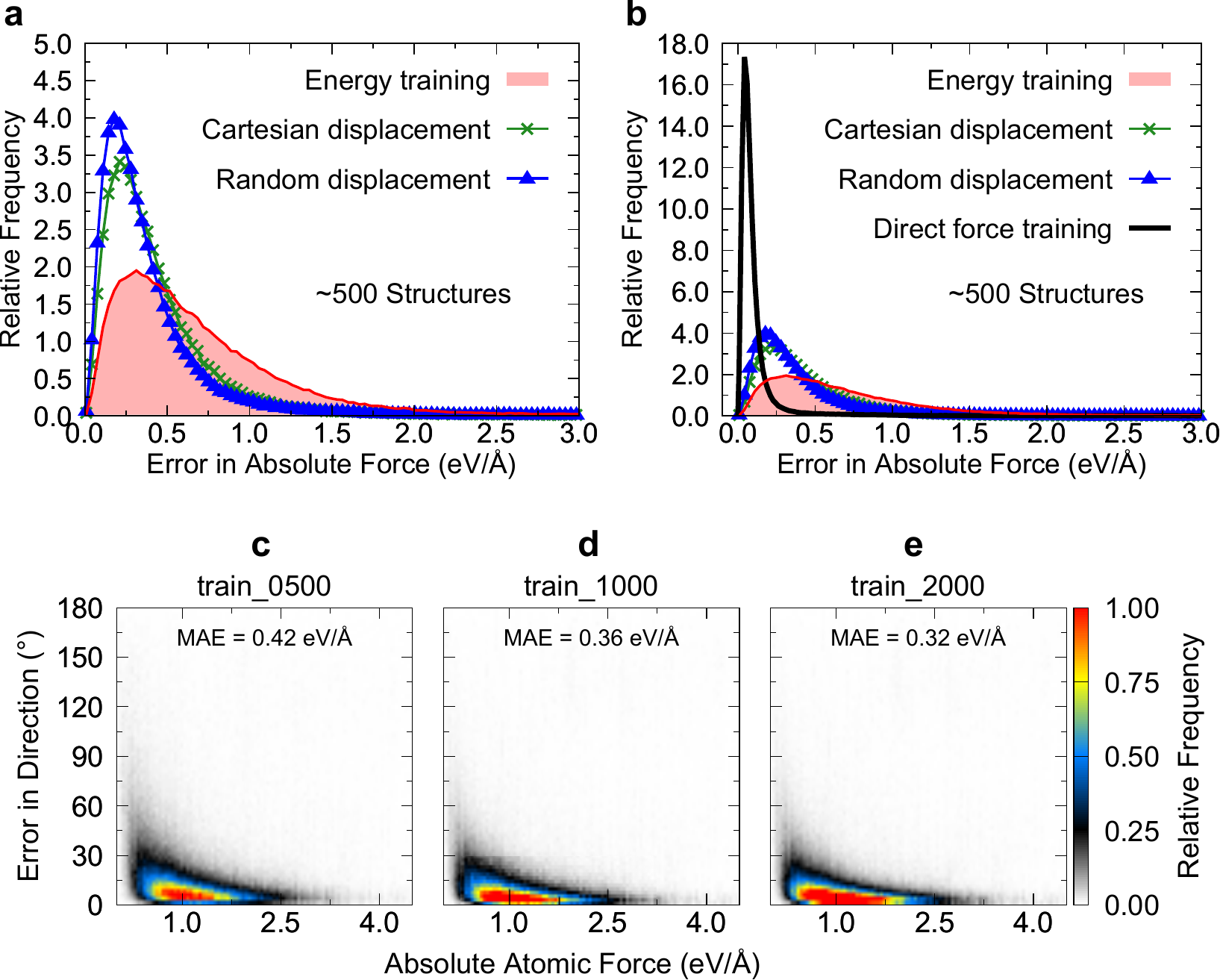}
  \caption{\label{fig:waterc-force-training}%
    \textbf{Impact of force training on the ANN force prediction errors
      for clusters of six water molecules.}  \textbf{a},~Force error
    distribution for training on the \texttt{train\_0500} training set
    only with energy information (\textit{Energy training}, red shade)
    and using approximate force training with the Taylor-expansion
    approach and the Cartesian (green crosses) and random (blue
    triangles) displacement strategies. \textbf{b}, The same data as in
    panel \textbf{a} but in addition the error distribution resulting
    from direct (exact) force training is shown as thick black line.
    \textbf{c-e},~Frequency of occurrence of a given error in the
    direction of the force in degrees as a function of the absolute
    atomic force, using approximate force training with the
    Taylor-expansion approach.  The optimal values for the displacement
    and the multiple $a$ were used (Cartesian displacement:
    $\delta = 0.03$~\AA{},\, $a=11$, Random displacement:
    $\delta_\textup{max} = 0.008$~\AA{},\, $a=22$).  All statistics
    shown are based on 10~ANN potentials trained for each scenario.}
\end{figure*}
Using the optimal parameters for the displacement $\delta$ and fraction
of generated structures $a$ for each of the three reference data sets
and both displacement strategies, we studied the errors in the
prediction of the atomic forces in order to compare both displacement
strategies with each other and to quantify the improvement of the
predicted forces with respect to the conventional approach where only
total energies of the reference structures are fitted.

\textbf{Figure~\ref{fig:waterc-force-training}a} compares the
distribution of absolute force errors in the validation set after
training on the \texttt{train\_0500} set with and without force
information.
As seen in the figure, the errors in the absolute force are drastically
reduced by both displacement strategies.
Especially the tail of the error distribution with errors greater than
1.0~eV/\AA{} has nearly disappeared for the ANN potentials that were
trained using the Taylor-expansion approach.
In addition, the prediction of the direction of the force vectors also
improves distinctly, as seen in
\textbf{Figure~\ref{fig:waterc-force-training}c-e}.
Comparing the results obtained with the Taylor-expansion approach to
those for conventional \hbox{energy training}
(\textbf{Figure~\ref{fig:trnset-size}b-d}), a general decrease of the
error in the predicted forces can be observed that is also reflected by
a reduction of the MAE.

Based on the error distributions in
\textbf{Figure~\ref{fig:waterc-force-training}a}, both displacement
strategies perform nearly equally well.
Displacing atoms along the Cartesian directions (strategy~A) generating
5,088 additional structures results in a slightly smaller improvement
than displacing atoms in random directions (strategy~B) generating
10,362 structures, but both Taylor-expansion strategies are a
significant improvement over energy-only training.
For the small water cluster system, the Cartesian displacement strategy
requires generating far fewer additional structures than the random
displacement approach.
We found that this is not generally the case and instead depends on the
number of atoms in the reference structures.
With increasing structure size, the optimal multiple $a$ increases more
rapidly for displacement strategy (A), and thus for larger structures
displacement strategy (B) becomes more efficient.

Comparison of \textbf{Figure~\ref{fig:waterc-force-training}a} with
\textbf{Figure~\ref{fig:trnset-size}a} and the heat map in
\textbf{Figure~\ref{fig:waterc-force-training}c} with that in
\textbf{Figure~\ref{fig:trnset-size}d} shows that training the
\texttt{train\_0500} data set with approximate force information results
in quantitatively similar error distributions as training on the
energies in the four-times larger \texttt{train\_2000} data set.
Hence, the approximate force training reduces the number of training
data points by a factor of four for the example of the water cluster
data set.

In principle, both the size of the reference data set as well as the
force training can be expected to affect also the error in the predicted
energy.
In the case of the water cluster data set, even training on the energies
of the \texttt{train\_0500} data set without force information results
in energy errors in the validation set that are well below
$2\times{}10^{-3}$~eV/atom, which is an order of magnitude below
chemical accuracy (1~kcal/mol~$\approx$~0.04~eV).
Therefore, the water cluster data set is not well suited to investigate
the impact of force training on the energy prediction, and we will
return to this question when discussing more complex data sets in the
following section~\ref{sec:water-bulk}.
For the water cluster data set, supplementary \textbf{Figure~S2} shows
that the distribution of the energy error is not significantly affected
by force training using the Taylor-expansion approach.

\subsubsection{Comparison of approximate and direct force training}

For the small water cluster structure (18~atoms) and the smallest
(\texttt{train\_0500}) data set, direct force training by including all
force errors in the loss function as described in
section~\ref{sec:force-training} is computationally feasible.
Note that even for this simple system, the direct force training took on
average \emph{eight times more computer time} per iteration (training
epoch) than the Taylor-expansion method with data multiple $a=22$ on our
computer system.
The force error distribution resulting from direct force training is
shown in \textbf{Figure~\ref{fig:waterc-force-training}b}.

Comparing the results obtained by approximate force training to those
from direct force training, it is obvious that direct force training
improves the force prediction even further.
The tail of the distribution of the absolute force error approaches zero
at around 0.5~eV/\AA{}, and as a result the error distribution is
significantly narrower than the ones obtained by the other methods with
its maximum being closer to~0.

The same trend is also found for the improvement of the prediction of
the direction of the force vector by direct force training compared to
the approximate Taylor-expansion approach.
The mean absolute force error for direct force training is 0.18
eV/\AA{}, and a heat map of the direction error is shown in
supplementary \textbf{Figure~S3}.

Quantitatively measured by the MAE, approximate force training using the
Taylor-expansion methodology gives around half the reduction of the
force error that is achieved by direct force training.
This can be most clearly seen in
\textbf{Figure~\ref{fig:displacement-and-fraction}}, in which the
relative MAE obtained after direct force training is indicated by a red
dashed line.

Even though direct force training leads to yet better predictions of the
forces, one has to take into account that it is significantly more
computationally demanding.
For larger and more complex systems for which direct force training
would be challenging or even infeasible, approximate force training with
the Taylor-expansion approach offers an alternative to improve the force
prediction significantly in comparison to conventional energy training.
Therefore, we investigate condensed phases in the following sections.

\subsection{Bulk water}
\label{sec:water-bulk}

We applied the Taylor-expansion methodology to a reference data set of a
periodic bulk water system with 64~water molecules (192~atoms) that was
generated by running an \emph{ab~initio} MD (AIMD) simulation at a
temperature of 400~K over a simulation time of 100~ps with time steps of
1~fs, producing a total of 100,000~MD frames.
The simulations employed a $\Gamma$-point k-point mesh for the numerical
Brillouin-zone integration.
All parameters of the DFT calculations are detailed in the methods
section~\ref{sec:methods}.

700~structures along the AIMD trajectory were selected for potential
training.
90\% of the reference data set were used for ANN potential training, and
the remaining 10\% were used as an independent test set to monitor
training progress, and to detect overfitting.
2,000 different equally-spaced MD frames from the complete trajectory
were compiled into a third independent set for validation, and all
results within this section are based on the validation set none of
which was used for training.

\begin{figure*}
  \linespread{1.0}
  \centering
  \includegraphics[width=0.8\textwidth]{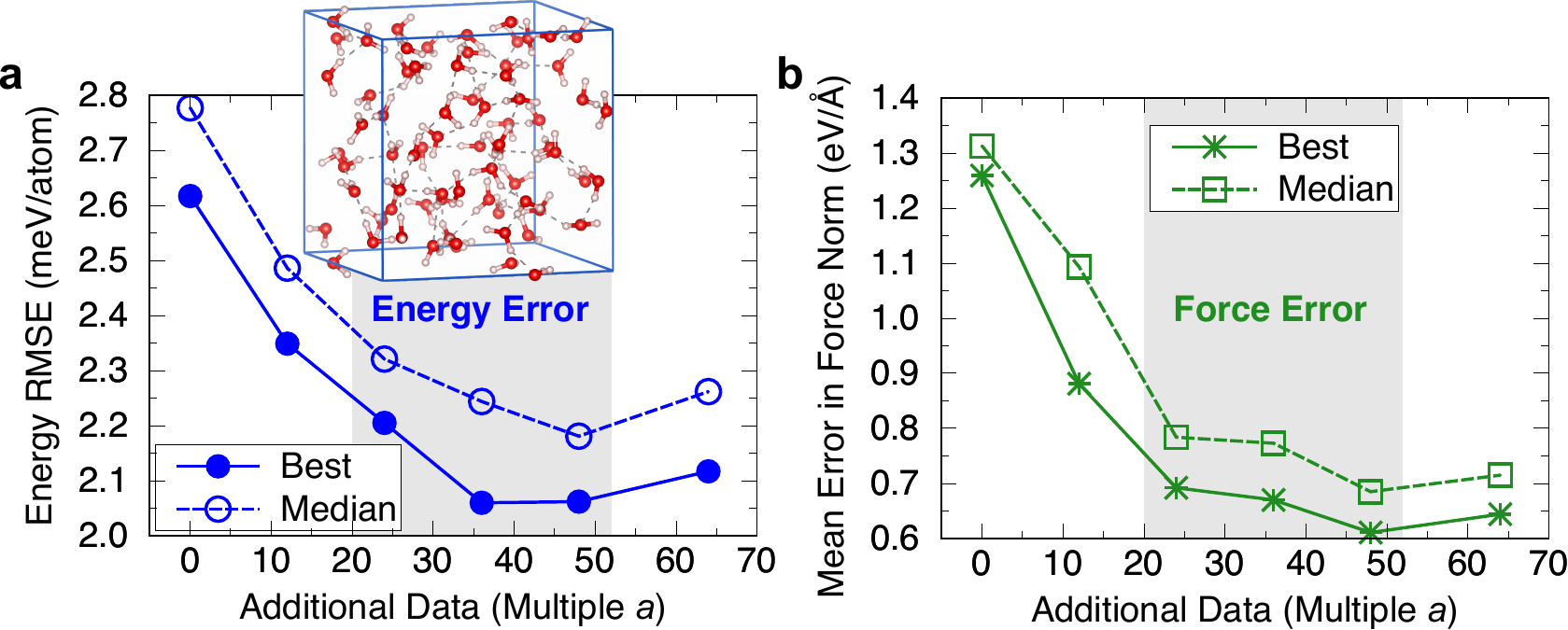}
  \caption{\label{fig:water-bulk}%
    \textbf{Change of the energy and force error with increasing number
      of additional structures from Taylor expansion for bulk water.}
    \textbf{a},~Root mean squared error (RMSE) of the energy relative to
    the DFT reference energies for ANN potentials trained with and
    without force information.
    The solid line shows the error of the best potential out of ten
    training runs, and the dashed line indicates the corresponding
    median. \textbf{b},~The equivalent analysis for the error in the
    norm of the interatomic forces.  A representative bulk water
    structure is shows as inset in panel \textbf{a} (oxygen atoms are
    red, and hydrogen is white).}
\end{figure*}
Since the nature of the bonds in the water bulk is the same as in the
water clusters of the previous section, a maximal displacement of
$\delta_{\textup{max}}=0.01$~\AA{} was used for the water bulk system,
which was found to be close to optimal for water clusters (see
Figure~\ref{fig:displacement-and-fraction}).
Additionally, we limit the discussion to random displacement
strategy~(B), since it performed better than displacing individual atoms
for water clusters.
Instead, here we focus on the impact of increasing the number of
additional structures generated by the Taylor-expansion approach.

In general, two competing effects can be expected: On the one hand, the
force training should become more effective with increasing multiple
$a$, i.e., with an increasing number of additional structures \emph{per
  original structure} generated by our approach.
On the other hand, the first-order Taylor-expansion is approximate, and
the energies of the additional structures are less accurate and noise is
introduced into the training set.
Thus, the accuracy of the energy fit could be expected to decrease after
introducing too many additional structures via atomic displacement.

To determine the balance of these two effects, data multiples between
$a=12$ and $a=64$ were considered, which corresponds to approximately
$12\times{}700=8,400$ to $64\times{}700=44,800$ additional structures
generated by Taylor expansion for the 700-structure reference data set.
See section~\ref{sec:ANN-parameters} for details of the ANN potential
construction.

\textbf{Figures~\ref{fig:water-bulk}a~and~b} show, as solid lines, the
best energy and force prediction errors out of ten ANN potentials
trained on the same data but with different initial weight parameters.
Additionally, the median ANN potential error is shown (dashed lines) as
a proxy for the likely result of a single ANN potential training run.
As seen in the figures, both the energy and force errors initially
decrease with increasing amount of additional data.
For small data multiples $a$, the force error improves significantly
from 1.26~eV/\AA{} (energy training only) to 0.88~eV/\AA{} (30\%
improvement) for $a=12$ and 0.69~eV/\AA{} (45\%) for $a=24$
(\textbf{Fig.~\ref{fig:water-bulk}b}).
Increasing $a$ beyond 24 only yields marginal further improvement, and
for $a=48$ the force error has decreased to 0.61~eV/\AA{} (52\%).

Interestingly, the energy RMSE decreases simultaneously from
2.6~meV/atom (energy training) to 2.1~meV/atom (19\%) for $a=36$
(\textbf{Fig.~\ref{fig:water-bulk}a}).
This improvement indicates that the displaced structures added to the
training set via Taylor expansion have improved the transferability of
the resulting ANN potential.
Increasing $a$ further does not result in further improvement of the
energy error, and an increase of the error is observed for $a=64$, which
is in line with our expectation that too many additional structures
introduce noise in the training set.

\begin{figure*}
  \linespread{1.0}
  \centering
  \includegraphics[width=0.8\textwidth]{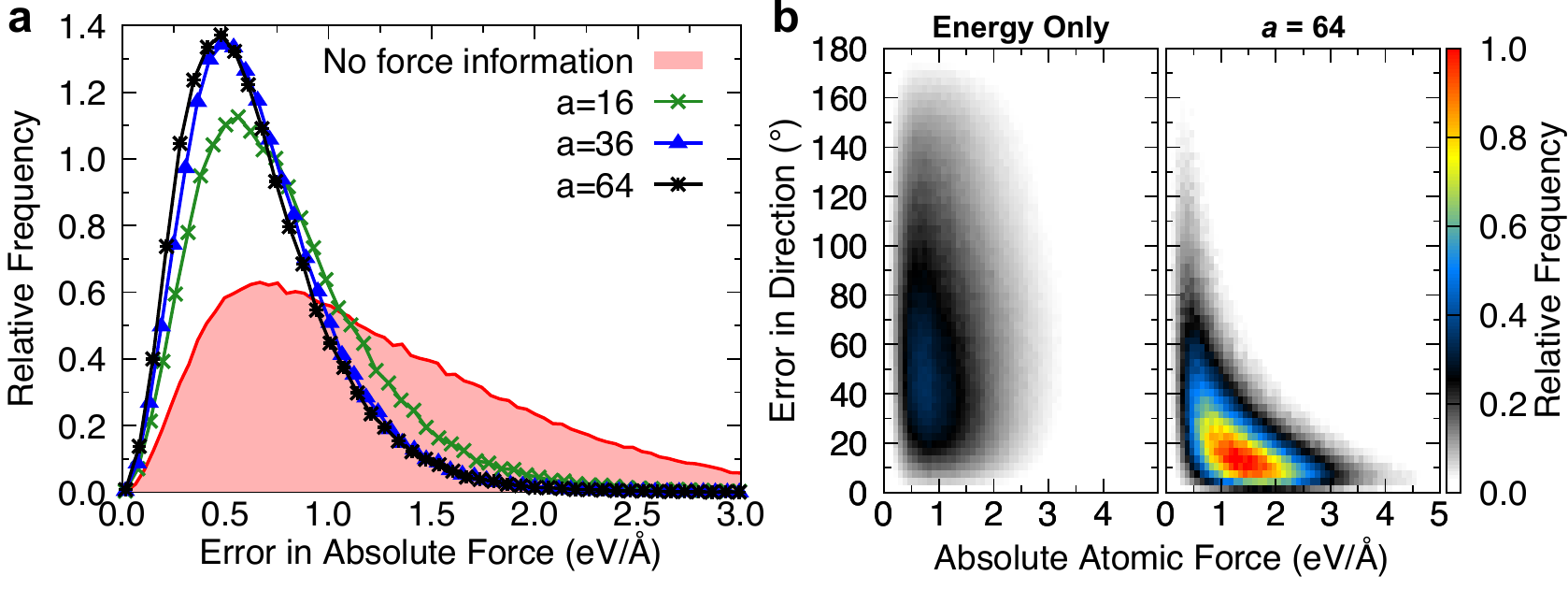}
  \caption{\label{fig:water-bulk-norm-angle}%
    \textbf{Error in the force norm and direction for bulk water.}
    \textbf{a},~Distribution of the absolute error of the atomic forces
    without force training (red area) and with force training using
    increasingly more additional structures generated by displacement.
    \textbf{b},~Distribution of the error in the force direction without
    (left) and with (right) force training.  For the force training, a
    maximal displacement of $\delta_{\textup{max}}=0.01$~\AA{} and a
    structure multiple of $a=64$ were used.}
\end{figure*}
Having quantified the general improvement of the predicted atomic forces
with the force-training approach, we analyze next where these
improvements originate from.
\textbf{Figure~\ref{fig:water-bulk-norm-angle}a} shows the distribution
of absolute force errors in the validation set with and without force
training.
In the case of energy training only, the force errors are widely spread
out, and for some atoms the force error is greater than 3.0~eV/\AA{}.
The largest atomic forces in the data set are around 5.0~eV/\AA{}, so
that 3.0~eV/\AA{} corresponds to an error of at least 60\%.
As discussed above, the force prediction error depends on the size of
the reference data set, and thus the analysis in
\textbf{Figure~\ref{fig:water-bulk-norm-angle}a} shows that the
700-structure data set is not sufficient for robust force prediction.

Force training with the Taylor-expansion approach results in a strong
improvement of the force prediction, and training with a data multiple
of $a=16$ already removes the high-error tail of the error distribution.
Increasing $a$ to~36 further reduces the width of the distribution
significantly.
The error distributions for a large data multiple of $a=64$ is nearly
identical to the distribution for $a=36$ and both are centered around
0.6~eV/\AA{}, showing that the force training has converged for this
data set in agreement with \textbf{Fig.~\ref{fig:water-bulk}}.

\textbf{Figure~\ref{fig:water-bulk-norm-angle}b} shows a heat map of the
distribution of the errors in the direction of the atomic forces for all
atoms in the validation set for training without forces (left) and with
forces (right).
The results for $a=64$ are shown in the figure, though $a=36$ yielded a
qualitatively equivalent error distribution.
As seen in the figure, when training only energies the errors in the
force direction are spread out over nearly the entire angle range.
In contrast, the force training improves the error distribution
significantly, yielding a maximum at around 15$^{\circ}$ and a strong
decay of the error with the norm of the force vectors.

\begin{figure}
  \linespread{1.0}
  \centering
  \includegraphics[width=\columnwidth]{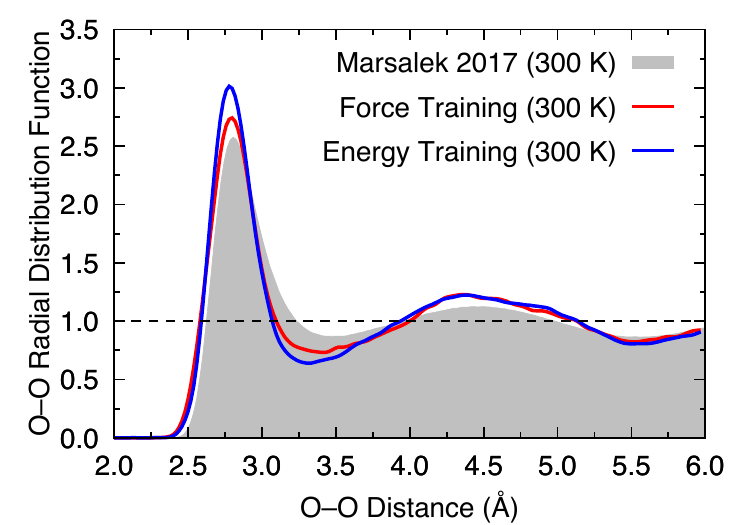}
  \caption{\label{fig:RDF}%
    \textbf{O–O radial distribution functions (RDF) for bulk water.} The
    RDF was calculated using preliminary ANN potentials trained with the
    Taylor-expansion approach (red line) and using energies only (blue
    line), respectively. The grey region indicates results from \textit{ab
    initio} MD simulations with the same density functional at 300~K from
    reference~\citenum{jpcl8-2017-1545}. The ANN-potential RDFs were
    evaluated for simulation cells with 128~water molecules.}
\end{figure}

As another test, we constructed preliminary ANN potentials trained with
and without force information on a larger reference data set of 10,000
frames taken from the same AIMD trajectory to evaluate the radial
distribution function (RDF) of liquid water.
In \textbf{Figure~\ref{fig:RDF}}, the results are compared to a literature
reference for an equivalent DFT approach.\cite{jpcl8-2017-1545}
Note that the construction of robust ANN potentials for water requires
data sets with very diverse structures and phases that include also
short O–H bond lengths that occur rarely in MD
simulations,\cite{pnas113-2016-8368, jpcl9-2018-851, tjopcl10-2019-6067}
and the preliminary potentials can only be considered a starting point
for the ANN potential construction and will require further refinement.

As seen in \textbf{Figure~\ref{fig:RDF}}, the ANN potential trained with
the Taylor-expansion approach yielded improvements in the RDF peak of
the first coordination shell.
Importantly, the stability of the MD simulations improved significantly
with force training.
Simulations using potentials trained on energies only were unstable due
to frequent extrapolation, whereas the potentials trained with force
information allowed for more robust MD simulations.
It is clear that the potential that was trained using forces via the
Taylor-expansion method performed better than the one that was trained
on energies only.




Direct force training of all atomic force components was not feasible
for the bulk water system.
As detailed in section~\ref{sec:force-training}, for condensed phases
direct force training scales with the number of atoms within twice the
potential cutoff radius.
For $R_{c}=6.5$~\AA{}, there are on average 960~atoms within range, so
that direct force training of all atomic forces would take
$\sim$1,000~times the computer time of energy-only training.

As shown above, the Taylor-expansion approach for force training works
robustly for both isolated water clusters and water bulk.
For materials with more complex compositions, sampling the structural
space with sufficient resolution is challenging, and force training
becomes even more important, as discussed in the next section.

\subsection{Quaternary metal oxide}
\label{sec:TM-oxide}

\vspace{-\baselineskip}

To determine the performance of the force-training approach for a
material with complex chemical composition we finally investigated a
quaternary transition metal oxide.
Our benchmark system, the \ce{Li-Mo-Ni-Ti} oxide (LMNTO), is of
technological relevance as prospective high-capacity positive electrode
material for lithium-ion batteries.~\cite{ees8-2015-3255}
The compound exhibits substitutional disorder in which all four metal
species, Li, Mo, Ni, and Ti, share the same sublattice.

\begin{figure*}
  \linespread{1.0}
  \centering
  \includegraphics[width=0.8\textwidth]{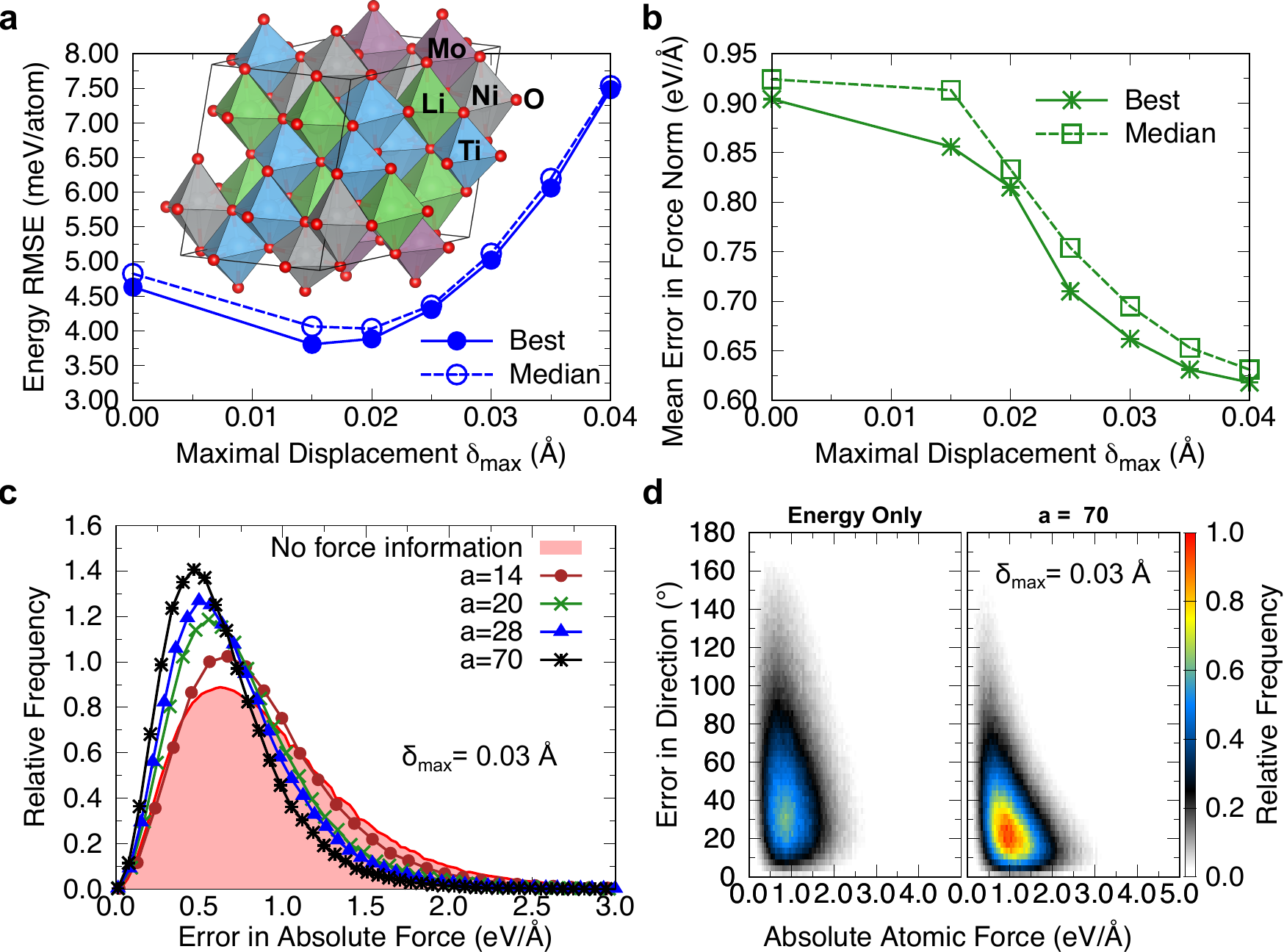}
  \caption{\label{fig:LNTMO}%
    \textbf{Approximate force training applied to a lithium
      transition-metal oxide.}  Change of \textbf{a}, the root mean
    squared error (RMSE) of the energy and \textbf{b}, the mean error in
    the force norm as function of the maximal displacement
    $\delta_{\textup{max}}$.  \textbf{c}, Distribution of the absolute
    force error for different training parameters.  \textbf{d},
    Distribution of the error in the force direction.  A representative
    crystal structure of the Li-Mo-Ni-Ti oxide is shown as inset of
    panel \textbf{a}.}
\end{figure*}

We generated a reference data set for LMNTO with composition
\ce{Li8Mo2Ni7Ti7O32} by running a 50~ps long AIMD simulation at 400~K as
described in detail in the methods section~\ref{sec:methods}, yielding a
total of 50,000~MD frames.
Again, training ($\sim$720~structures), test ($\sim$80~structures), and
validation (1,800~structures) sets were generated.
A representative structure of the LMNTO unit cell is shown as inset in
\textbf{Fig.~\ref{fig:LNTMO}a}.

The bonding in lithium transition-metal oxides exhibits mostly ionic
character, and the bond strength cannot be expected to be the same as in
the water systems of the previous sections.
Therefore, the maximal displacement $\delta_{\textup{max}}$ for the
Taylor-expansion approach first needs to be optimized for LMNTO.

\textbf{Figure~\ref{fig:LNTMO}a and b} show the change of the energy and
force errors as function of $\delta_{\textup{max}}$ for a fixed
additional data multiple of $a=70$.
As seen in subfigure~\textbf{b}, the mean error in the predicted force
decreases with increasing maximal displacement and has not yet converged
for $\delta_{\textup{max}}=0.04$~\AA{}, which is four times greater than the
optimal displacement found for water.
On the other hand, the median energy RMSE has a minimum at around
$\delta_{\textup{max}}=0.02$~\AA{} beyond which the energy error
increases.
A maximal displacement of $\delta_{\textup{max}}=0.03$~\AA{} is a good
compromise for LMNTO, resulting in a small decrease in accuracy for the
energy from 4.6~meV/atom to 4.9~meV/atom (increase by 6.5\%) but a
strong improvement in the accuracy of force prediction from
0.92~eV/\AA{} to 0.66~eV/\AA{} (28.3\%).

The distribution of errors in the absolute values and directions of the
atomic forces are shown in \textbf{Figs.~\ref{fig:LNTMO}c and d},
respectively.
The results for $\delta_{\textup{max}}=0.03$~\AA{} are shown.
Qualitatively, the same improvement of the absolute force error as in
the case of the water systems is seen, though even without force
training the error distribution does not exhibit any high-error tail.
The force training also results in an improved accuracy in the direction
of the predicted forces, though the maximum of the error distribution is
at a slightly higher value of around $\sim$20$^{\circ}$.

\subsubsection{Towards application in MD simulations of LMNTO}

\begin{figure*}
  \linespread{1.0}
  \centering
  \includegraphics[width=0.7\textwidth]{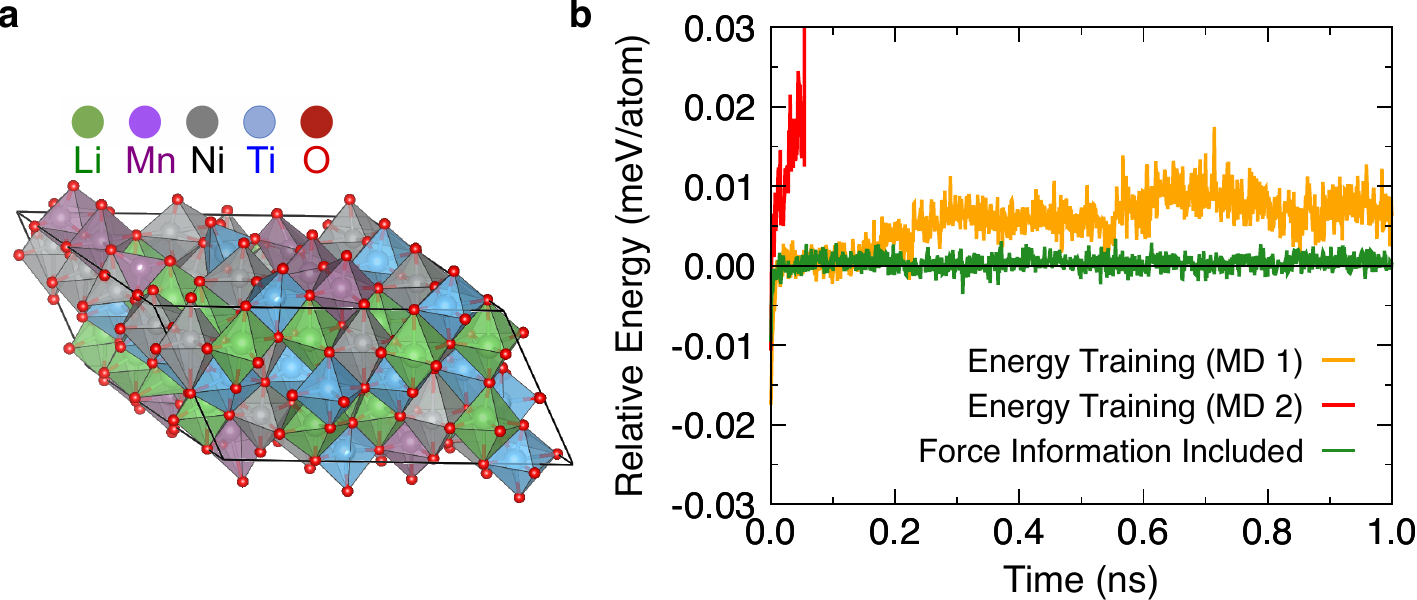}
  \caption{\label{fig:LNTMO-MD}%
    \textbf{Energy conservation in microcanonical molecular dynamics
      (MD).}  \textbf{a},~Representative LMNTO structure taken from MD
    simulation (colors as in Figure~\ref{fig:LNTMO}). \textbf{b},~Change
    of the total energy during MD simulations in the microcanonical
    ($NVE$) statistical ensemble using ANN potentials trained on energy
    information only (orange and red lines) and trained using the
    Taylor-expansion approach (green line).}
\end{figure*}
The MAE of the forces and the RMSE of the energy are abstract quality
measures of the ANN potential, but in practice the robustness and
reliability of the potential in an actual application is most important.
To construct a preliminary ANN potential for LMNTO, we compiled a data
set of $\sim$4,000~AIMD frames that were used as reference for training
three ANN potentials with force information
($\delta_{\textup{max}}=0.015$~\AA{}, $a=20$) and three without force
information.
The construction of accurate ANN potentials for materials with complex
structures and compositions typically requires training sets with tens
of thousands of reference structures.~\cite{pssb250-2013-1191,
  nl14-2014-2670, arxiv1901-2019-09272}
4,000~reference structures taken from a single AIMD trajectory can
function as an initial data set for the construction of a preliminary
ANN potential as a starting point for subsequent iterative
refinement.~\cite{acie56-2017-12828, cms114-2016-135}

The robustness of an interatomic potential and the smoothness of the
predicted PES is reflected by the numerical energy conservation in MD
simulations.
To test energy conservation, we carried out MD simulations in the
microcanonical ($NVE$) statistical ensemble using a time step of 1.0~fs,
and the resulting change of the total energy over the course of 1.0~ns
is shown in \textbf{Figure~\ref{fig:LNTMO-MD}} for different ANN
potentials.
The MD simulations were performed for a $2\times{}2\times{}1$~LMNTO
supercell containing 224~atoms (\textbf{Figure~\ref{fig:LNTMO-MD}a})
that was previously thermally equilibrated at a temperature of 400~K
with AIMD simulation.
All of the ANN potentials trained with approximate force information
conserved the total energy well with fluctuations on the order of
$10^{-3}$~meV/atom (one example is shown in
\textbf{Figure~\ref{fig:LNTMO-MD}b}).
However, those potentials that were trained only on the energies showed
numerical instabilities of varying degree, and two examples are plotted
in \textbf{Figure~\ref{fig:LNTMO-MD}b}.
\vfill\null

\newpage
This experiment demonstrates that the additional structure-energy data
points generated from force information using the Taylor-expansion
approach contribute to the transferability of the ANN potential and
improve the smoothness of the potential interpolation.


\section{Discussion}
\label{sec:discussion}

We introduced a computationally efficient method for the simultaneous
training of energies and interatomic forces for the construction of
accurate ANN potentials.
We assessed the methodology for three relevant complex materials
systems: water cluster structures, bulk liquid water, and a solid metal
oxide.
We demonstrated that the approach increases the accuracy of predicted
forces, reduces the number of first principles reference data points
needed, and can improve the transferability of ANN potentials.

In general, we find that force training has the greatest impact for
small reference data sets.
With force training, a small reference data set of $\sim$500~water
cluster structures could achieve nearly the same predictive power as
training only the energy on a reference data set with
$\sim$2,000~structures.
Using the force-training approach, the structural and chemical space can
thus be more coarsely sampled than using energy-only training.
This has important implications for the construction of ANN potentials
for materials with complex compositions, such as the quaternary metal
oxide of section~\ref{sec:TM-oxide}, for which an exhaustive sampling of
the structural and chemical space is infeasible.

The Taylor-expansion methodology is approximate, and it is meant as a
computationally more efficient alternative to the direct training of
force information with a modified loss function.
Direct force training also incurs a significant memory overhead if the
derivatives of the atomic-structure descriptors are stored in memory,
whereas the Taylor-expansion approach uses exactly the same amount of
memory as training with energy information only.
As demonstrated for the water cluster data set, direct force training
can further improve the force prediction when it is computationally
feasible.
However, even for the small water cluster system, direct force training
was already eight times more computationally demanding than approximate
force training, and owing to its formal quadratic scaling this
difference will be even greater for larger systems.

We discussed two different strategies for the generation of additional
approximate reference energy data points from force information by
approximating the energy of structures with slightly displaced atoms
using a first-order Taylor expansion.
The two strategies differ in the number of atoms that are displaced, but
both rely on randomization, either for the selection of atoms or to
decide the direction and magnitude of atomic displacements.
Displacing all atoms in a reference structure by small random vectors
with maximal length $\delta_{\textup{max}}$ is shown to be a robust
scheme for the generation of derived structures.
However, this strategy does, in principle, not rule out that multiple
additional structures with similar information content (i.e., similar
displacements) are generated.
The generated atomic displacements are not necessarily linearly
independent.

The atomic forces of a structure with $N$ atoms comprise $3N$ pieces of
information from the three force components of all atoms.
Hence, at most $3N$ structures with independent information content can
be derived from any given structure with energy and force information.
One set of generated structures with linearly independent atomic
displacements is given by the set of $3N$ \emph{normal
  modes}.~\cite{sm108-2015-1, prl119-2017-176402}
Excluding the 3~translations and 3~rotations that do not change the
energy, one could employ displacements in the remaining $(3N-6)$ normal
mode directions as a third strategy for our force training approach.
Normal-mode sampling has previously been shown to be a useful strategy
for the generation of reference data sets, especially for molecular
systems.\cite{cs8-2017-3192, nc10-2019-2903}
However, we found empirically that far fewer than $3N$ additional
structures are required to converge the error of the interatomic forces
for a given reference data set.
For the bulk water system with 64~molecules, $N=3\times{}64=192$, so
that the number of normal modes is $3\times{}192-6=570$, but our
computational experiments show that the force error already plateaus for
24--48~additional structures with random atomic displacements.
An exhaustive enumeration of all degrees of freedom would therefore be
sub-optimal for our force-training methodology, and it is not obvious
which normal modes should be selected if a subset was to be used.
The random displacement strategy offers a reasonable compromise of
information density and generality.

As evident from the water clusters of section~\ref{sec:water-clusters}
and the oxide system of section~\ref{sec:TM-oxide} the optimal maximal
displacement $\delta_{\textup{max}}$ for use in the Taylor extrapolation
is system dependent.
The energy change with atomic displacement depends on the
material-specific bond strength, i.e., the force constants of the
interatomic bonds.
The covalent O-H bonds in water are more rigid than the ionic bonds in
LMNTO, so that smaller displacements are needed for water than for the
oxide.
As a rule of thumb, we expect the optimal value of
$\delta_{\textup{max}}$ to be approximately proportional to the smallest
interatomic distances (bond lengths) in the reference data set, which is
$\sim$0.9~\AA{} for the \ce{O-H} bond in water and $\sim$1.7~\AA{} for
\ce{O-metal} bonds in LMNTO, though the nature of the bonds (covalent,
ionic, metallic, or dispersive) will also have an impact.

A related point is the dependence of the energy error on the value of
$\delta_{\textup{max}}$.
As discussed for the oxide system and shown in
\textbf{Figs.~\ref{fig:LNTMO}a and b}, the optimal displacement can be a
compromise of force and energy accuracy.
If the displacement is chosen too large, it is no longer a good
approximation that the PES is linear, and the error of the first-order
Taylor expansion becomes too large.
For each materials system, it is thus necessary to benchmark different
displacement values to determine the value that is optimal for both
force and energy training.

It is important to note that the Taylor expansion extrapolation approach
does not have to be limited to first order.
If higher derivatives of the potential energy are available, e.g., if
the Hessian matrix has been calculated for the reference structures,
then higher-order Taylor expansions can be used to improve the accuracy
of the energy extrapolation.

The methodology could be further extended and refined for specific
applications.
Atomic displacement could be limited to atoms with high force components
to avoid introducing noise in shallow regions of the PES.
Structures from geometry optimizations are generally not useful for our
approach as the atomic forces are near zero, and such structures should
not be considered for displacement.
It could also be useful to allow selection of specific atoms for
displacement, so that the force prediction accuracy can be increased for
select substructures.
Similarly, the maximal displacement $\delta_{\textup{max}}$ could be
made a species-specific parameter, which would be especially useful in
interface systems containing domains with different types of bonding,
such as solid-liquid interfaces.
These directions will be explored in future work.

\section{Conclusions}
\label{sec:conclusions}

We introduced a new computationally efficient method for the training of
accurate artificial neural network (ANN) potentials on interatomic force
information and established its effectiveness for different classes of
materials.
The methodology is based on a Taylor extrapolation of the total energy
using the atomic forces from the reference calculations without the need
for additional electronic structure calculations.
Training occurs on approximate energies, and the computationally
demanding evaluation of the second derivatives of the ANN function is
not required.
Translating the force information to approximate energies makes it
possible to bypass the quadratic scaling with the number of atoms that
conventional force-training methods exhibit, so that the
Taylor-expansion approach can be used for reference data sets containing
complex atomic structures.
We showed that approximate force training can improve the force
prediction error by around 50\% of direct force training, and is
computationally efficient even for systems for which the latter is
challenging or infeasible.
We demonstrated for three example systems, a cluster of six water
molecules, liquid water, and a complex metal oxide, that the
force-training approach
\textit{(i)}~allows to substantially reduce the size of the reference
data sets for ANN potential construction, by nearly 75\% in the case of
the water cluster data set;
\textit{(ii)}~increases the transferability of the ANN potential by
improving the energy prediction accuracy for unseen structures; and
\textit{(iii)}~generally improves the force prediction accuracy, leading
to improved stability of MD simulations.
The alternative force training approach simplifies the construction of
general ANN potentials for the prediction of accurate energies and
interatomic forces and is in principle applicable to any type of
material.


\section{Methods}
\label{sec:methods}

\subsection{Electronic structure calculations}
\label{sec:DFT}

\paragraph{Water clusters}

The reference data set was generated by performing ab initio molecular
dynamics (MD) simulations with DL\_POLY \cite{jmc16-2006-1911} via
Chemshell.\cite{jms632-2003-1, wcms4-2014-101}
The reference data set was generated in an iterative manner.  First
three MD simulation runs were performed, where the semiemperical GFN-xTB
\cite{jctc13-2017-1989,jctc15-2019-1652} was used as ab initio method.
All of these simulations were run for 30~ps with a time step of 0.5~fs.
The initial velocities were chosen randomly according to a
Maxwell-Boltzmann distribution. The temperature simulated was 300~K for
the first two simulations, the third MD simulation run was performed at
800~K.
The temperature was in all cases controlled by a Nosé-Hoover
thermostat.~\cite{tjocp81-1984-511, pra31-1985-1695}
During the simulations, a harmonic restraint was applied on all atoms to
keep the cluster of water molecules confined to a sphere with a radius
of about 5~\AA{}.
For the restraint a harmonic force with force constant
190.5~eV 
was applied to any atom that has a distance greater than 0.0005~\AA{} to
the 1st atom of the structure (central atom).
In order to get a realistic approximation of the energy for the
structures obtained from the MD simulation runs, the energies and forces
for all structures were recalculated using the BLYP-D3 functional
\cite{pra38-1988-3098, prb37-1988-785, jcp98-1993-5612,
  jcp101-1994-7729, jcp132-2010-154104, jpcl7-2016-2197} with the basis
set def2-TZVP.
For the single point energy calculations Turbomole\cite{ircms4-2014-91}
was used via Chemshell.\cite{jms632-2003-1, wcms4-2014-101}
Using this reference data set, two neural networks were trained to
obtain a first approximation of the ANN potential.
On each of the obtained ANN potentials a MD simulation at 300~K was
performed for 75~ps.
The other parameters for the MD simulation were chosen as discussed
before.
From these MD simulations on the ANN potentials 4420 additional
reference structures were obtained. As for the AIMD reference structures
the energies and forces were recalculated on BLYP-D3/def2-TZVP level of
theory.

\paragraph{Periodic AIMD simulations}
The periodic AIMD simulations were carried out with the Vienna Ab initio
simulation package (VASP)\cite{prB54-1996-11169, cms6-1996-15} and
projector-augmented wave (PAW) pseudopotentials.\cite{prB50-1994-17953}
For the bulk water system the revised Perdew-Burke-Ernzerhof density
functional~\cite{prl80-1998-890} with the Grimme D3 van-der-Waals
correction~\cite{jcp132-2010-154104} (revPBE+D3) was used that has
previously been shown to be reliable for water.~\cite{pnas113-2016-8368}
The AIMD simulations of the Li-Mo-Ni-Ti-O system employed the strongly
constrained and appropriately normed (SCAN) semilocal density
functional.\cite{prl115-2015-036402}

For both periodic systems, the plane-wave cutoff was 400~eV, and
$\Gamma$-point only k-point meshes were employed.
A time step of 1~fs was used for the integration of the equation of
motion, and a Nosé-Hoover thermostat\cite{tjocp81-1984-511,
  pra31-1985-1695} was used to maintain the temperature at 400~K.
The bulk water data set was compiled by collecting every 100th frame
from the first 70~ps of the AIMD trajectory, yielding a reference data
set of 700~structures.

\subsection{Reference data and ANN potential training}
\label{sec:ANN-parameters}

The Taylor-expansion approach for force training was implemented in the
\emph{atomic energy network} (ænet) package,\cite{cms114-2016-135} which
was used for the construction and application of the reported ANN
potentials.
In the present work, the limited memory BFGS
algorithm\cite{siam16-1995-1190, toms23-1997-550} was used for the ANN
weight optimization (training).
The Artrith--Urban--Ceder Chebyshev descriptor for local atomic
environments\cite{prb96-2017-014112} was employed if not otherwise
noted.
Further details for the different materials systems follow.

\paragraph{Lennard-Jones dimer}
For the dimer data set, a Chebyshev descriptor with radial expansion
order 10 was employed.
No angular expansion was used since the dimer does not have any bond
angles.
The ANN was comprised of two hidden layers with each five nodes and
hyperbolic tangent activation, i.e., the ANN architecture was
11--5--5--1.

\paragraph{Water clusters}
For the water clusters symmetry functions by Behler and Parrinello
\cite{prl98-2007-146401,jcp134-2011-074106} were used as descriptor.
The parameters for the symmetry functions for water were taken from the
publication of T.Morawietz et al. \cite{pnas113-2016-8368}.
For the training of the ANN the reference set was divided randomly into
a training and a test set. Thereby 90\% of the structures were used as
the training set and the remaining 10\% of the structures were used as
the test set which was used to measure the quality of the predictions
obtained from the ANN potential for structures that have not been used
in the fit of the potential.
The ANN architecture $N_\text{symm}$--10--10--1 was used, where
$N_\text{symm}$ is given by the descriptor \cite{pnas113-2016-8368} with
$N_\text{symm}=27$ for hydrogen and $N_\text{symm}=30$ for oxygen.  The
hyperbolic tangent was used as activation function.

Each ANN was trained for 5000 epochs, then the weights and biases for
the epoch that that lead to the smallest test set error during training
were used.

\paragraph{Water bulk}
For the bulk water ANN potential, a Chebyshev descriptor with a radial
expansion order of~18 and an angular expansion order of 4~was used.
The interaction cutoffs were 6.5~\AA{} and 3.0~\AA{} for the radial and
angular expansion, respectively, and the ANN architecture was
48--10--10--1, i.e., two hidden layers with each ten nodes were used.
This corresponds to a total of 611~weight parameters.

\paragraph{\ce{Li-Ni-Ti-Mo-O} system}
A descriptor with an interaction range of 6.5~\AA{} and an expansion
order of 20 for the radial distribution function and a range of
3.5~\AA{} and an expansion order of 2~for the angular interactions was
used.
For the 700-structure data set, the best balance of model complexity and
accuracy was obtained for an ANN potential with 2 hidden layers and each
10 nodes (48-10-10-1).

\subsection{ANN potential MD simulations}

All ANN potential MD simulations were carried out using the Tinker
software~\cite{jocc8-1987-1016} and ANN potentials via an interface with
the ænet package and used the Verlet algorithm\cite{pr159-67-98.pdf} for
the integration of the equation of motion.

The ANN potentials for the water bulk MD simulations were trained on a
data set of $\sim$10,000 frames taken from an AIMD trajectory at 400~K
and were based on a 48-20-20-1 ANN architecture.
The radial distribution functions (RDF) shown in Figure~\ref{fig:RDF}
were evaluated for a simulation cell with 128~water molecules using MD
simulations in the canonical ($NVT$) ensemble by averaging over a total
of 10~ps after a 2~ps equilibration period.
A time step of 0.25~fs was used.
A Bussi–Parrinello thermostat~\cite{jcp126-2007-14101} was employed for
the NVT sampling, and the target temperature was 300~K.

For the MD simulations in Figure~\ref{fig:LNTMO-MD}, a larger
\ce{Li-Ni-Ti-Mo-O} reference data set with $\sim$4,000~structure was
used for training.
For this data set, we employed a 48-20-20-1 ANN architecture.
A time-step of 1~fs was used.

\section*{Data availability}

The reference data sets of the water cluster, bulk water, and
transition-metal oxide systems can be obtained from the Materials Cloud
repository
(\href{https://doi.org/10.24435/materialscloud:2020.0037/v1}{doi: 10.24435/materialscloud:2020.0037/v1}).
These data sets contain atomic structures and interatomic forces in the
XCrySDen\cite{jmgm17-1999-176} structure format (XSF), and total
energies are included as additional meta information.

\section*{Code availability}

This work made use of the free and open-source atomic energy network
(ænet) package.
The ænet source code can be obtained either from the ænet website
(\url{http://ann.atomistic.net}) or from GitHub
(\url{https://github.com/atomisticnet/aenet}).
The Taylor-expansion approach will be made available in the next release
of the ænet package.


\section{Acknowledgments}
\label{sec:acknowledgments}

N.A.\ and A.U.\ acknowledge financial support by the U.S. Department of
Energy (DOE) Office of Energy Efficiency and Renewable Energy, Vehicle
Technologies Office, Contract No.~DE-SC0012704.
A.C.\ and J.K.\ acknowledge financial support by the European Union's
Horizon 2020 research and innovation programme (Grant Agreement
No.~646717, TUNNELCHEM) and the German Research Foundation (DFG) through
the Cluster of Excellence in Simulation Technology (No.~EXC 310/2) at
the University of Stuttgart.
Computational resources for the water cluster DFT calculations were
provided by the state of Baden-W\"urttemberg through bwHPC and the
German Research Foundation (DFG) through grant no INST 40/467-1 FUGG.
Development of ænet used the Extreme Science and Engineering Discovery
Environment (XSEDE), which is supported by National Science Foundation
grant number ACI-1053575.
The LMNTO DFT calculations used resources of the Center for Functional
Nanomaterials, which is a U.S. DOE Office of Science Facility, at
Brookhaven National Laboratory under Contract No.~DE-SC0012704.
For the simulations of bulk water, we acknowledge computing resources
from Columbia University's Shared Research Computing Facility project,
which is supported by NIH Research Facility Improvement Grant
1G20RR030893-01, and associated funds from the New York State Empire
State Development, Division of Science Technology and Innovation
(NYSTAR) Contract C090171, both awarded April 15, 2010.
N.A.\ and A.U.\ thank Dr.~Penghao Xiao for valuable discussions.
We thank Dr.~Tobias Morawietz for testing our implementation and for
insightful discussions.

\section{Contributions}
\label{sec:contributions}

N.A.~designed and planned the study.
N.A.\ and A.U.\ supervised all aspects of the project.
A.C., A.U., and N.A.\ implemented the methodology into ænet, and
performed extensive tests.
A.C.\ and N.A.\ performed DFT calculations and constructed the ANN
potentials.
A.C., A.U., and N.A.\ analyzed the results.
The manuscript was written by A.C., A.U., and N.A.\ with the help of
J.K.
All authors contributed to discussions and commented on the manuscript.


\bibliographystyle{aipnum4-1}
\bibliography{Force-Training-Bibliography.bib}

\begin{thebibliography}{77}%
\makeatletter
\providecommand \@ifxundefined [1]{%
 \@ifx{#1\undefined}
}%
\providecommand \@ifnum [1]{%
 \ifnum #1\expandafter \@firstoftwo
 \else \expandafter \@secondoftwo
 \fi
}%
\providecommand \@ifx [1]{%
 \ifx #1\expandafter \@firstoftwo
 \else \expandafter \@secondoftwo
 \fi
}%
\providecommand \natexlab [1]{#1}%
\providecommand \enquote  [1]{``#1''}%
\providecommand \bibnamefont  [1]{#1}%
\providecommand \bibfnamefont [1]{#1}%
\providecommand \citenamefont [1]{#1}%
\providecommand \href@noop [0]{\@secondoftwo}%
\providecommand \href [0]{\begingroup \@sanitize@url \@href}%
\providecommand \@href[1]{\@@startlink{#1}\@@href}%
\providecommand \@@href[1]{\endgroup#1\@@endlink}%
\providecommand \@sanitize@url [0]{\catcode `\\12\catcode `\$12\catcode
  `\&12\catcode `\#12\catcode `\^12\catcode `\_12\catcode `\%12\relax}%
\providecommand \@@startlink[1]{}%
\providecommand \@@endlink[0]{}%
\providecommand \url  [0]{\begingroup\@sanitize@url \@url }%
\providecommand \@url [1]{\endgroup\@href {#1}{\urlprefix }}%
\providecommand \urlprefix  [0]{URL }%
\providecommand \Eprint [0]{\href }%
\providecommand \doibase [0]{http://dx.doi.org/}%
\providecommand \selectlanguage [0]{\@gobble}%
\providecommand \bibinfo  [0]{\@secondoftwo}%
\providecommand \bibfield  [0]{\@secondoftwo}%
\providecommand \translation [1]{[#1]}%
\providecommand \BibitemOpen [0]{}%
\providecommand \bibitemStop [0]{}%
\providecommand \bibitemNoStop [0]{.\EOS\space}%
\providecommand \EOS [0]{\spacefactor3000\relax}%
\providecommand \BibitemShut  [1]{\csname bibitem#1\endcsname}%
\let\auto@bib@innerbib\@empty
\bibitem [{\citenamefont {Behler}\ and\ \citenamefont
  {Parrinello}(2007)}]{prl98-2007-146401}%
  \BibitemOpen
  \bibfield  {author} {\bibinfo {author} {\bibfnamefont {J.}~\bibnamefont
  {Behler}}\ and\ \bibinfo {author} {\bibfnamefont {M.}~\bibnamefont
  {Parrinello}},\ }\href {\doibase 10.1103/PhysRevLett.98.146401} {\bibfield
  {journal} {\bibinfo  {journal} {Phys. Rev. Lett.}\ }\textbf {\bibinfo
  {volume} {98}},\ \bibinfo {pages} {146401} (\bibinfo {year}
  {2007})}\BibitemShut {NoStop}%
\bibitem [{\citenamefont {Bart\'ok}\ \emph {et~al.}(2010)\citenamefont
  {Bart\'ok}, \citenamefont {Payne}, \citenamefont {Kondor},\ and\
  \citenamefont {Cs\'anyi}}]{prl104-2010-136403}%
  \BibitemOpen
  \bibfield  {author} {\bibinfo {author} {\bibfnamefont {A.~P.}\ \bibnamefont
  {Bart\'ok}}, \bibinfo {author} {\bibfnamefont {M.~C.}\ \bibnamefont {Payne}},
  \bibinfo {author} {\bibfnamefont {R.}~\bibnamefont {Kondor}}, \ and\ \bibinfo
  {author} {\bibfnamefont {G.}~\bibnamefont {Cs\'anyi}},\ }\href {\doibase
  10.1103/PhysRevLett.104.136403} {\bibfield  {journal} {\bibinfo  {journal}
  {Phys. Rev. Lett.}\ }\textbf {\bibinfo {volume} {104}},\ \bibinfo {pages}
  {136403} (\bibinfo {year} {2010})}\BibitemShut {NoStop}%
\bibitem [{\citenamefont {Thompson}\ \emph {et~al.}(2014)\citenamefont
  {Thompson}, \citenamefont {Swiler}, \citenamefont {Trott}, \citenamefont
  {Foiles},\ and\ \citenamefont {Tucker}}]{jocp-2014-SNAP}%
  \BibitemOpen
  \bibfield  {author} {\bibinfo {author} {\bibfnamefont {A.}~\bibnamefont
  {Thompson}}, \bibinfo {author} {\bibfnamefont {L.}~\bibnamefont {Swiler}},
  \bibinfo {author} {\bibfnamefont {C.}~\bibnamefont {Trott}}, \bibinfo
  {author} {\bibfnamefont {S.}~\bibnamefont {Foiles}}, \ and\ \bibinfo {author}
  {\bibfnamefont {G.}~\bibnamefont {Tucker}},\ }\href {\doibase
  10.1016/j.jcp.2014.12.018} {\bibfield  {journal} {\bibinfo  {journal} {J.
  Comput. Phys.}\ }\textbf {\bibinfo {volume} {285}},\ \bibinfo {pages} {316}
  (\bibinfo {year} {2014})}\BibitemShut {NoStop}%
\bibitem [{\citenamefont {Bart\'{o}k}\ and\ \citenamefont
  {Cs\'{a}nyi}(2015)}]{ijqc115-2015-1051}%
  \BibitemOpen
  \bibfield  {author} {\bibinfo {author} {\bibfnamefont {A.~P.}\ \bibnamefont
  {Bart\'{o}k}}\ and\ \bibinfo {author} {\bibfnamefont {G.}~\bibnamefont
  {Cs\'{a}nyi}},\ }\href {\doibase 10.1002/qua.24927} {\bibfield  {journal}
  {\bibinfo  {journal} {Int. J. Quantum Chem.}\ }\textbf {\bibinfo {volume}
  {115}},\ \bibinfo {pages} {1051} (\bibinfo {year} {2015})}\BibitemShut
  {NoStop}%
\bibitem [{\citenamefont {Behler}(2017)}]{acie56-2017-12828}%
  \BibitemOpen
  \bibfield  {author} {\bibinfo {author} {\bibfnamefont {J.}~\bibnamefont
  {Behler}},\ }\href {\doibase 10.1002/anie.201703114} {\bibfield  {journal}
  {\bibinfo  {journal} {Angew. Chem. Int. Ed.}\ }\textbf {\bibinfo {volume}
  {56}},\ \bibinfo {pages} {12828} (\bibinfo {year} {2017})}\BibitemShut
  {NoStop}%
\bibitem [{\citenamefont {Zong}\ \emph {et~al.}(2018)\citenamefont {Zong},
  \citenamefont {Pilania}, \citenamefont {Ding}, \citenamefont {Ackland},\ and\
  \citenamefont {Lookman}}]{ncm4-2018-48}%
  \BibitemOpen
  \bibfield  {author} {\bibinfo {author} {\bibfnamefont {H.}~\bibnamefont
  {Zong}}, \bibinfo {author} {\bibfnamefont {G.}~\bibnamefont {Pilania}},
  \bibinfo {author} {\bibfnamefont {X.}~\bibnamefont {Ding}}, \bibinfo {author}
  {\bibfnamefont {G.~J.}\ \bibnamefont {Ackland}}, \ and\ \bibinfo {author}
  {\bibfnamefont {T.}~\bibnamefont {Lookman}},\ }\href {\doibase
  10.1038/s41524-018-0103-x} {\bibfield  {journal} {\bibinfo  {journal} {npj
  Comput. Mater.}\ }\textbf {\bibinfo {volume} {4}},\ \bibinfo {pages} {48}
  (\bibinfo {year} {2018})}\BibitemShut {NoStop}%
\bibitem [{\citenamefont {Himanen}\ \emph {et~al.}(2019)\citenamefont
  {Himanen}, \citenamefont {Geurts}, \citenamefont {Foster},\ and\
  \citenamefont {Rinke}}]{as-2019-1900808}%
  \BibitemOpen
  \bibfield  {author} {\bibinfo {author} {\bibfnamefont {L.}~\bibnamefont
  {Himanen}}, \bibinfo {author} {\bibfnamefont {A.}~\bibnamefont {Geurts}},
  \bibinfo {author} {\bibfnamefont {A.~S.}\ \bibnamefont {Foster}}, \ and\
  \bibinfo {author} {\bibfnamefont {P.}~\bibnamefont {Rinke}},\ }\href
  {\doibase 10.1002/advs.201900808} {\bibfield  {journal} {\bibinfo  {journal}
  {Adv. Sci.}\ }\textbf {\bibinfo {volume} {6}},\ \bibinfo {pages} {1900808}
  (\bibinfo {year} {2019})}\BibitemShut {NoStop}%
\bibitem [{\citenamefont {Nyshadham}\ \emph {et~al.}(2019)\citenamefont
  {Nyshadham}, \citenamefont {Rupp}, \citenamefont {Bekker}, \citenamefont
  {Shapeev}, \citenamefont {Mueller}, \citenamefont {Rosenbrock}, \citenamefont
  {Cs{\'{a}}nyi}, \citenamefont {Wingate},\ and\ \citenamefont
  {Hart}}]{ncm5-2019-51}%
  \BibitemOpen
  \bibfield  {author} {\bibinfo {author} {\bibfnamefont {C.}~\bibnamefont
  {Nyshadham}}, \bibinfo {author} {\bibfnamefont {M.}~\bibnamefont {Rupp}},
  \bibinfo {author} {\bibfnamefont {B.}~\bibnamefont {Bekker}}, \bibinfo
  {author} {\bibfnamefont {A.~V.}\ \bibnamefont {Shapeev}}, \bibinfo {author}
  {\bibfnamefont {T.}~\bibnamefont {Mueller}}, \bibinfo {author} {\bibfnamefont
  {C.~W.}\ \bibnamefont {Rosenbrock}}, \bibinfo {author} {\bibfnamefont
  {G.}~\bibnamefont {Cs{\'{a}}nyi}}, \bibinfo {author} {\bibfnamefont {D.~W.}\
  \bibnamefont {Wingate}}, \ and\ \bibinfo {author} {\bibfnamefont {G.~L.~W.}\
  \bibnamefont {Hart}},\ }\href {\doibase 10.1038/s41524-019-0189-9} {\bibfield
   {journal} {\bibinfo  {journal} {npj Comput. Mater.}\ }\textbf {\bibinfo
  {volume} {5}},\ \bibinfo {pages} {51} (\bibinfo {year} {2019})}\BibitemShut
  {NoStop}%
\bibitem [{\citenamefont {Kostiuchenko}\ \emph {et~al.}(2019)\citenamefont
  {Kostiuchenko}, \citenamefont {Körmann}, \citenamefont {Neugebauer},\ and\
  \citenamefont {Shapeev}}]{ncm5-2019-55}%
  \BibitemOpen
  \bibfield  {author} {\bibinfo {author} {\bibfnamefont {T.}~\bibnamefont
  {Kostiuchenko}}, \bibinfo {author} {\bibfnamefont {F.}~\bibnamefont
  {Körmann}}, \bibinfo {author} {\bibfnamefont {J.}~\bibnamefont
  {Neugebauer}}, \ and\ \bibinfo {author} {\bibfnamefont {A.}~\bibnamefont
  {Shapeev}},\ }\href {\doibase 10.1038/s41524-019-0195-y} {\bibfield
  {journal} {\bibinfo  {journal} {npj Comput. Mater.}\ }\textbf {\bibinfo
  {volume} {5}},\ \bibinfo {pages} {55} (\bibinfo {year} {2019})}\BibitemShut
  {NoStop}%
\bibitem [{\citenamefont {Deng}\ \emph {et~al.}(2019)\citenamefont {Deng},
  \citenamefont {Chen}, \citenamefont {Li},\ and\ \citenamefont
  {Ong}}]{ncm5-2019-75}%
  \BibitemOpen
  \bibfield  {author} {\bibinfo {author} {\bibfnamefont {Z.}~\bibnamefont
  {Deng}}, \bibinfo {author} {\bibfnamefont {C.}~\bibnamefont {Chen}}, \bibinfo
  {author} {\bibfnamefont {X.-G.}\ \bibnamefont {Li}}, \ and\ \bibinfo {author}
  {\bibfnamefont {S.~P.}\ \bibnamefont {Ong}},\ }\href {\doibase
  10.1038/s41524-019-0212-1} {\bibfield  {journal} {\bibinfo  {journal} {npj
  Comput. Mater.}\ }\textbf {\bibinfo {volume} {5}},\ \bibinfo {pages} {75}
  (\bibinfo {year} {2019})}\BibitemShut {NoStop}%
\bibitem [{\citenamefont {Schmidt}\ \emph {et~al.}(2019)\citenamefont
  {Schmidt}, \citenamefont {Marques}, \citenamefont {Botti},\ and\
  \citenamefont {Marques}}]{ncm5-2019-83}%
  \BibitemOpen
  \bibfield  {author} {\bibinfo {author} {\bibfnamefont {J.}~\bibnamefont
  {Schmidt}}, \bibinfo {author} {\bibfnamefont {M.~R.~G.}\ \bibnamefont
  {Marques}}, \bibinfo {author} {\bibfnamefont {S.}~\bibnamefont {Botti}}, \
  and\ \bibinfo {author} {\bibfnamefont {M.~A.~L.}\ \bibnamefont {Marques}},\
  }\href {\doibase 10.1038/s41524-019-0221-0} {\bibfield  {journal} {\bibinfo
  {journal} {npj Comput. Mater.}\ }\textbf {\bibinfo {volume} {5}},\ \bibinfo
  {pages} {83} (\bibinfo {year} {2019})}\BibitemShut {NoStop}%
\bibitem [{\citenamefont {Hohenberg}\ and\ \citenamefont
  {Kohn}(1964)}]{pr136-64-846}%
  \BibitemOpen
  \bibfield  {author} {\bibinfo {author} {\bibfnamefont {P.}~\bibnamefont
  {Hohenberg}}\ and\ \bibinfo {author} {\bibfnamefont {W.}~\bibnamefont
  {Kohn}},\ }\href {\doibase 10.1103/PhysRev.136.B864} {\bibfield  {journal}
  {\bibinfo  {journal} {Phys. Rev.}\ }\textbf {\bibinfo {volume} {136}},\
  \bibinfo {pages} {B864} (\bibinfo {year} {1964})}\BibitemShut {NoStop}%
\bibitem [{\citenamefont {Kohn}\ and\ \citenamefont
  {Sham}(1965)}]{pr140-65-A1133}%
  \BibitemOpen
  \bibfield  {author} {\bibinfo {author} {\bibfnamefont {W.}~\bibnamefont
  {Kohn}}\ and\ \bibinfo {author} {\bibfnamefont {L.~J.}\ \bibnamefont
  {Sham}},\ }\href {\doibase 10.1103/PhysRev.140.A1133} {\bibfield  {journal}
  {\bibinfo  {journal} {Phys. Rev.}\ }\textbf {\bibinfo {volume} {140}},\
  \bibinfo {pages} {A1133} (\bibinfo {year} {1965})}\BibitemShut {NoStop}%
\bibitem [{\citenamefont {Burke}(2012)}]{jcp136-2012-150901}%
  \BibitemOpen
  \bibfield  {author} {\bibinfo {author} {\bibfnamefont {K.}~\bibnamefont
  {Burke}},\ }\href {\doibase 10.1063/1.4704546} {\bibfield  {journal}
  {\bibinfo  {journal} {J. Chem. Phys.}\ }\textbf {\bibinfo {volume} {136}},\
  \bibinfo {pages} {150901} (\bibinfo {year} {2012})}\BibitemShut {NoStop}%
\bibitem [{\citenamefont {Jones}(2015)}]{rmp87-2015-897}%
  \BibitemOpen
  \bibfield  {author} {\bibinfo {author} {\bibfnamefont {R.~O.}\ \bibnamefont
  {Jones}},\ }\href {\doibase 10.1103/revmodphys.87.897} {\bibfield  {journal}
  {\bibinfo  {journal} {Rev. Mod. Phys.}\ }\textbf {\bibinfo {volume} {87}},\
  \bibinfo {pages} {897} (\bibinfo {year} {2015})}\BibitemShut {NoStop}%
\bibitem [{\citenamefont {Cooper}\ and\ \citenamefont
  {Kästner}(2019)}]{jpca123-2019-9061}%
  \BibitemOpen
  \bibfield  {author} {\bibinfo {author} {\bibfnamefont {A.~M.}\ \bibnamefont
  {Cooper}}\ and\ \bibinfo {author} {\bibfnamefont {J.}~\bibnamefont
  {Kästner}},\ }\href {\doibase 10.1021/acs.jpca.9b07013} {\bibfield
  {journal} {\bibinfo  {journal} {J. Phys. Chem. A}\ }\textbf {\bibinfo
  {volume} {123}},\ \bibinfo {pages} {9061} (\bibinfo {year}
  {2019})}\BibitemShut {NoStop}%
\bibitem [{\citenamefont {Smith}\ \emph {et~al.}(2019)\citenamefont {Smith},
  \citenamefont {Nebgen}, \citenamefont {Zubatyuk}, \citenamefont {Lubbers},
  \citenamefont {Devereux}, \citenamefont {Barros}, \citenamefont {Tretiak},
  \citenamefont {Isayev},\ and\ \citenamefont {Roitberg}}]{nc10-2019-2903}%
  \BibitemOpen
  \bibfield  {author} {\bibinfo {author} {\bibfnamefont {J.~S.}\ \bibnamefont
  {Smith}}, \bibinfo {author} {\bibfnamefont {B.~T.}\ \bibnamefont {Nebgen}},
  \bibinfo {author} {\bibfnamefont {R.}~\bibnamefont {Zubatyuk}}, \bibinfo
  {author} {\bibfnamefont {N.}~\bibnamefont {Lubbers}}, \bibinfo {author}
  {\bibfnamefont {C.}~\bibnamefont {Devereux}}, \bibinfo {author}
  {\bibfnamefont {K.}~\bibnamefont {Barros}}, \bibinfo {author} {\bibfnamefont
  {S.}~\bibnamefont {Tretiak}}, \bibinfo {author} {\bibfnamefont
  {O.}~\bibnamefont {Isayev}}, \ and\ \bibinfo {author} {\bibfnamefont {A.~E.}\
  \bibnamefont {Roitberg}},\ }\href {\doibase 10.1038/s41467-019-10827-4}
  {\bibfield  {journal} {\bibinfo  {journal} {Nat. Commun.}\ }\textbf {\bibinfo
  {volume} {10}},\ \bibinfo {pages} {2903} (\bibinfo {year}
  {2019})}\BibitemShut {NoStop}%
\bibitem [{\citenamefont {Schran}, \citenamefont {Behler},\ and\ \citenamefont
  {Marx}(2019)}]{jctc16-2019-88}%
  \BibitemOpen
  \bibfield  {author} {\bibinfo {author} {\bibfnamefont {C.}~\bibnamefont
  {Schran}}, \bibinfo {author} {\bibfnamefont {J.}~\bibnamefont {Behler}}, \
  and\ \bibinfo {author} {\bibfnamefont {D.}~\bibnamefont {Marx}},\ }\href
  {\doibase 10.1021/acs.jctc.9b00805} {\bibfield  {journal} {\bibinfo
  {journal} {J. Chem. Theory Comput.}\ }\textbf {\bibinfo {volume} {16}},\
  \bibinfo {pages} {88} (\bibinfo {year} {2019})}\BibitemShut {NoStop}%
\bibitem [{\citenamefont {Artrith}, \citenamefont {Morawietz},\ and\
  \citenamefont {Behler}(2011)}]{prB83-2011-153101}%
  \BibitemOpen
  \bibfield  {author} {\bibinfo {author} {\bibfnamefont {N.}~\bibnamefont
  {Artrith}}, \bibinfo {author} {\bibfnamefont {T.}~\bibnamefont {Morawietz}},
  \ and\ \bibinfo {author} {\bibfnamefont {J.}~\bibnamefont {Behler}},\ }\href
  {\doibase 10.1103/PhysRevB.83.153101} {\bibfield  {journal} {\bibinfo
  {journal} {Phys. Rev. B}\ }\textbf {\bibinfo {volume} {83}},\ \bibinfo
  {pages} {153101} (\bibinfo {year} {2011})}\BibitemShut {NoStop}%
\bibitem [{\citenamefont {Artrith}, \citenamefont {Urban},\ and\ \citenamefont
  {Ceder}(2017)}]{prb96-2017-014112}%
  \BibitemOpen
  \bibfield  {author} {\bibinfo {author} {\bibfnamefont {N.}~\bibnamefont
  {Artrith}}, \bibinfo {author} {\bibfnamefont {A.}~\bibnamefont {Urban}}, \
  and\ \bibinfo {author} {\bibfnamefont {G.}~\bibnamefont {Ceder}},\ }\href
  {\doibase 10.1103/physrevb.96.014112} {\bibfield  {journal} {\bibinfo
  {journal} {Phys. Rev. B}\ }\textbf {\bibinfo {volume} {96}},\ \bibinfo
  {pages} {014112} (\bibinfo {year} {2017})}\BibitemShut {NoStop}%
\bibitem [{\citenamefont {Artrith}\ and\ \citenamefont
  {Behler}(2012)}]{prB85-2012-045439}%
  \BibitemOpen
  \bibfield  {author} {\bibinfo {author} {\bibfnamefont {N.}~\bibnamefont
  {Artrith}}\ and\ \bibinfo {author} {\bibfnamefont {J.}~\bibnamefont
  {Behler}},\ }\href {\doibase 10.1103/PhysRevB.85.045439} {\bibfield
  {journal} {\bibinfo  {journal} {Phys. Rev. B}\ }\textbf {\bibinfo {volume}
  {85}},\ \bibinfo {pages} {045439} (\bibinfo {year} {2012})}\BibitemShut
  {NoStop}%
\bibitem [{\citenamefont {Sun}\ and\ \citenamefont
  {Sautet}(2018)}]{jotacs140-2018-2812}%
  \BibitemOpen
  \bibfield  {author} {\bibinfo {author} {\bibfnamefont {G.}~\bibnamefont
  {Sun}}\ and\ \bibinfo {author} {\bibfnamefont {P.}~\bibnamefont {Sautet}},\
  }\href {\doibase 10.1021/jacs.7b11239} {\bibfield  {journal} {\bibinfo
  {journal} {J. Am. Chem. Soc.}\ }\textbf {\bibinfo {volume} {140}},\ \bibinfo
  {pages} {2812} (\bibinfo {year} {2018})}\BibitemShut {NoStop}%
\bibitem [{\citenamefont {Sosso}\ \emph {et~al.}(2012)\citenamefont {Sosso},
  \citenamefont {Miceli}, \citenamefont {Caravati}, \citenamefont {Behler},\
  and\ \citenamefont {Bernasconi}}]{prb85-2012-174103}%
  \BibitemOpen
  \bibfield  {author} {\bibinfo {author} {\bibfnamefont {G.~C.}\ \bibnamefont
  {Sosso}}, \bibinfo {author} {\bibfnamefont {G.}~\bibnamefont {Miceli}},
  \bibinfo {author} {\bibfnamefont {S.}~\bibnamefont {Caravati}}, \bibinfo
  {author} {\bibfnamefont {J.}~\bibnamefont {Behler}}, \ and\ \bibinfo {author}
  {\bibfnamefont {M.}~\bibnamefont {Bernasconi}},\ }\href {\doibase
  10.1103/physrevb.85.174103} {\bibfield  {journal} {\bibinfo  {journal} {Phys.
  Rev. B}\ }\textbf {\bibinfo {volume} {85}},\ \bibinfo {pages} {174103}
  (\bibinfo {year} {2012})}\BibitemShut {NoStop}%
\bibitem [{\citenamefont {Artrith}\ and\ \citenamefont
  {Kolpak}(2014)}]{nl14-2014-2670}%
  \BibitemOpen
  \bibfield  {author} {\bibinfo {author} {\bibfnamefont {N.}~\bibnamefont
  {Artrith}}\ and\ \bibinfo {author} {\bibfnamefont {A.~M.}\ \bibnamefont
  {Kolpak}},\ }\href {\doibase 10.1021/nl5005674} {\bibfield  {journal}
  {\bibinfo  {journal} {Nano Lett.}\ }\textbf {\bibinfo {volume} {14}},\
  \bibinfo {pages} {2670} (\bibinfo {year} {2014})}\BibitemShut {NoStop}%
\bibitem [{\citenamefont {Elias}\ \emph {et~al.}(2016)\citenamefont {Elias},
  \citenamefont {Artrith}, \citenamefont {Bugnet}, \citenamefont {Giordano},
  \citenamefont {Botton}, \citenamefont {Kolpak},\ and\ \citenamefont
  {Shao-Horn}}]{acsc6-2016-1675}%
  \BibitemOpen
  \bibfield  {author} {\bibinfo {author} {\bibfnamefont {J.~S.}\ \bibnamefont
  {Elias}}, \bibinfo {author} {\bibfnamefont {N.}~\bibnamefont {Artrith}},
  \bibinfo {author} {\bibfnamefont {M.}~\bibnamefont {Bugnet}}, \bibinfo
  {author} {\bibfnamefont {L.}~\bibnamefont {Giordano}}, \bibinfo {author}
  {\bibfnamefont {G.~A.}\ \bibnamefont {Botton}}, \bibinfo {author}
  {\bibfnamefont {A.~M.}\ \bibnamefont {Kolpak}}, \ and\ \bibinfo {author}
  {\bibfnamefont {Y.}~\bibnamefont {Shao-Horn}},\ }\href {\doibase
  10.1021/acscatal.5b02666} {\bibfield  {journal} {\bibinfo  {journal} {{ACS}
  Catalysis}\ }\textbf {\bibinfo {volume} {6}},\ \bibinfo {pages} {1675}
  (\bibinfo {year} {2016})}\BibitemShut {NoStop}%
\bibitem [{\citenamefont {Morawietz}\ \emph {et~al.}(2016)\citenamefont
  {Morawietz}, \citenamefont {Singraber}, \citenamefont {Dellago},\ and\
  \citenamefont {Behler}}]{pnas113-2016-8368}%
  \BibitemOpen
  \bibfield  {author} {\bibinfo {author} {\bibfnamefont {T.}~\bibnamefont
  {Morawietz}}, \bibinfo {author} {\bibfnamefont {A.}~\bibnamefont
  {Singraber}}, \bibinfo {author} {\bibfnamefont {C.}~\bibnamefont {Dellago}},
  \ and\ \bibinfo {author} {\bibfnamefont {J.}~\bibnamefont {Behler}},\ }\href
  {\doibase 10.1073/pnas.1602375113} {\bibfield  {journal} {\bibinfo  {journal}
  {Proc. Natl. Acad. Sci. U.S.A.}\ }\textbf {\bibinfo {volume} {113}},\
  \bibinfo {pages} {8368} (\bibinfo {year} {2016})}\BibitemShut {NoStop}%
\bibitem [{\citenamefont {Smith}, \citenamefont {Isayev},\ and\ \citenamefont
  {Roitberg}(2017)}]{cs8-2017-3192}%
  \BibitemOpen
  \bibfield  {author} {\bibinfo {author} {\bibfnamefont {J.~S.}\ \bibnamefont
  {Smith}}, \bibinfo {author} {\bibfnamefont {O.}~\bibnamefont {Isayev}}, \
  and\ \bibinfo {author} {\bibfnamefont {A.~E.}\ \bibnamefont {Roitberg}},\
  }\href {\doibase 10.1039/c6sc05720a} {\bibfield  {journal} {\bibinfo
  {journal} {Chem. Sci.}\ }\textbf {\bibinfo {volume} {8}},\ \bibinfo {pages}
  {3192} (\bibinfo {year} {2017})}\BibitemShut {NoStop}%
\bibitem [{\citenamefont {Cooper}, \citenamefont {Hallmen},\ and\ \citenamefont
  {Kästner}(2018)}]{jcp148-2018-094106}%
  \BibitemOpen
  \bibfield  {author} {\bibinfo {author} {\bibfnamefont {A.~M.}\ \bibnamefont
  {Cooper}}, \bibinfo {author} {\bibfnamefont {P.~P.}\ \bibnamefont {Hallmen}},
  \ and\ \bibinfo {author} {\bibfnamefont {J.}~\bibnamefont {Kästner}},\
  }\href {\doibase 10.1063/1.5015950} {\bibfield  {journal} {\bibinfo
  {journal} {J. Chem. Phys.}\ }\textbf {\bibinfo {volume} {148}},\ \bibinfo
  {pages} {094106} (\bibinfo {year} {2018})}\BibitemShut {NoStop}%
\bibitem [{\citenamefont {Morawietz}\ \emph {et~al.}(2019)\citenamefont
  {Morawietz}, \citenamefont {Urbina}, \citenamefont {Wise}, \citenamefont
  {Wu}, \citenamefont {Lu}, \citenamefont {Ben-Amotz},\ and\ \citenamefont
  {Markland}}]{tjopcl10-2019-6067}%
  \BibitemOpen
  \bibfield  {author} {\bibinfo {author} {\bibfnamefont {T.}~\bibnamefont
  {Morawietz}}, \bibinfo {author} {\bibfnamefont {A.~S.}\ \bibnamefont
  {Urbina}}, \bibinfo {author} {\bibfnamefont {P.~K.}\ \bibnamefont {Wise}},
  \bibinfo {author} {\bibfnamefont {X.}~\bibnamefont {Wu}}, \bibinfo {author}
  {\bibfnamefont {W.}~\bibnamefont {Lu}}, \bibinfo {author} {\bibfnamefont
  {D.}~\bibnamefont {Ben-Amotz}}, \ and\ \bibinfo {author} {\bibfnamefont
  {T.~E.}\ \bibnamefont {Markland}},\ }\href {\doibase
  10.1021/acs.jpclett.9b01781} {\bibfield  {journal} {\bibinfo  {journal} {J.
  Phys. Chem. Lett.}\ }\textbf {\bibinfo {volume} {10}},\ \bibinfo {pages}
  {6067} (\bibinfo {year} {2019})}\BibitemShut {NoStop}%
\bibitem [{\citenamefont {Li}\ \emph {et~al.}(2017)\citenamefont {Li},
  \citenamefont {Ando}, \citenamefont {Minamitani},\ and\ \citenamefont
  {Watanabe}}]{tjocp147-2017-214106}%
  \BibitemOpen
  \bibfield  {author} {\bibinfo {author} {\bibfnamefont {W.}~\bibnamefont
  {Li}}, \bibinfo {author} {\bibfnamefont {Y.}~\bibnamefont {Ando}}, \bibinfo
  {author} {\bibfnamefont {E.}~\bibnamefont {Minamitani}}, \ and\ \bibinfo
  {author} {\bibfnamefont {S.}~\bibnamefont {Watanabe}},\ }\href {\doibase
  10.1063/1.4997242} {\bibfield  {journal} {\bibinfo  {journal} {J. Chem.
  Phys.}\ }\textbf {\bibinfo {volume} {147}},\ \bibinfo {pages} {214106}
  (\bibinfo {year} {2017})}\BibitemShut {NoStop}%
\bibitem [{\citenamefont {Artrith}, \citenamefont {Urban},\ and\ \citenamefont
  {Ceder}(2018)}]{jcp148-2018-241711}%
  \BibitemOpen
  \bibfield  {author} {\bibinfo {author} {\bibfnamefont {N.}~\bibnamefont
  {Artrith}}, \bibinfo {author} {\bibfnamefont {A.}~\bibnamefont {Urban}}, \
  and\ \bibinfo {author} {\bibfnamefont {G.}~\bibnamefont {Ceder}},\ }\href
  {\doibase 10.1063/1.5017661} {\bibfield  {journal} {\bibinfo  {journal} {J.
  Chem. Phys.}\ }\textbf {\bibinfo {volume} {148}},\ \bibinfo {pages} {241711}
  (\bibinfo {year} {2018})}\BibitemShut {NoStop}%
\bibitem [{\citenamefont {Lacivita}, \citenamefont {Artrith},\ and\
  \citenamefont {Ceder}(2018)}]{cm30-2019-7077}%
  \BibitemOpen
  \bibfield  {author} {\bibinfo {author} {\bibfnamefont {V.}~\bibnamefont
  {Lacivita}}, \bibinfo {author} {\bibfnamefont {N.}~\bibnamefont {Artrith}}, \
  and\ \bibinfo {author} {\bibfnamefont {G.}~\bibnamefont {Ceder}},\ }\href
  {\doibase 10.1021/acs.chemmater.8b02812} {\bibfield  {journal} {\bibinfo
  {journal} {Chem. Mater.}\ }\textbf {\bibinfo {volume} {30}},\ \bibinfo
  {pages} {7077} (\bibinfo {year} {2018})}\BibitemShut {NoStop}%
\bibitem [{\citenamefont {Artrith}\ \emph {et~al.}(2019)\citenamefont
  {Artrith}, \citenamefont {Urban}, \citenamefont {Wang},\ and\ \citenamefont
  {Ceder}}]{arxiv1901-2019-09272}%
  \BibitemOpen
  \bibfield  {author} {\bibinfo {author} {\bibfnamefont {N.}~\bibnamefont
  {Artrith}}, \bibinfo {author} {\bibfnamefont {A.}~\bibnamefont {Urban}},
  \bibinfo {author} {\bibfnamefont {Y.}~\bibnamefont {Wang}}, \ and\ \bibinfo
  {author} {\bibfnamefont {G.}~\bibnamefont {Ceder}},\ }\href@noop {}
  {\bibfield  {journal} {\bibinfo  {journal} {arXiv}\ }\textbf {\bibinfo
  {volume} {https://arxiv.org/abs/1901.09272}} (\bibinfo {year} {2019})},\
  \Eprint {http://arxiv.org/abs/1901.09272} {1901.09272} \BibitemShut {NoStop}%
\bibitem [{\citenamefont {Artrith}, \citenamefont {Hiller},\ and\ \citenamefont
  {Behler}(2013)}]{pssb250-2013-1191}%
  \BibitemOpen
  \bibfield  {author} {\bibinfo {author} {\bibfnamefont {N.}~\bibnamefont
  {Artrith}}, \bibinfo {author} {\bibfnamefont {B.}~\bibnamefont {Hiller}}, \
  and\ \bibinfo {author} {\bibfnamefont {J.}~\bibnamefont {Behler}},\ }\href
  {\doibase 10.1002/pssb.201248370} {\bibfield  {journal} {\bibinfo  {journal}
  {physica status solidi (b)}\ }\textbf {\bibinfo {volume} {250}},\ \bibinfo
  {pages} {1191} (\bibinfo {year} {2013})}\BibitemShut {NoStop}%
\bibitem [{\citenamefont {Natarajan}\ and\ \citenamefont
  {Behler}(2016)}]{pccp18-2016-28704}%
  \BibitemOpen
  \bibfield  {author} {\bibinfo {author} {\bibfnamefont {S.~K.}\ \bibnamefont
  {Natarajan}}\ and\ \bibinfo {author} {\bibfnamefont {J.}~\bibnamefont
  {Behler}},\ }\href {\doibase 10.1039/c6cp05711j} {\bibfield  {journal}
  {\bibinfo  {journal} {Phys. Chem. Chem. Phys.}\ }\textbf {\bibinfo {volume}
  {18}},\ \bibinfo {pages} {28704} (\bibinfo {year} {2016})}\BibitemShut
  {NoStop}%
\bibitem [{\citenamefont {Quaranta}\ \emph {et~al.}(2018)\citenamefont
  {Quaranta}, \citenamefont {Hellström}, \citenamefont {Behler}, \citenamefont
  {Kullgren}, \citenamefont {Mitev},\ and\ \citenamefont
  {Hermansson}}]{tjocp148-2018-241720}%
  \BibitemOpen
  \bibfield  {author} {\bibinfo {author} {\bibfnamefont {V.}~\bibnamefont
  {Quaranta}}, \bibinfo {author} {\bibfnamefont {M.}~\bibnamefont
  {Hellström}}, \bibinfo {author} {\bibfnamefont {J.}~\bibnamefont {Behler}},
  \bibinfo {author} {\bibfnamefont {J.}~\bibnamefont {Kullgren}}, \bibinfo
  {author} {\bibfnamefont {P.~D.}\ \bibnamefont {Mitev}}, \ and\ \bibinfo
  {author} {\bibfnamefont {K.}~\bibnamefont {Hermansson}},\ }\href {\doibase
  10.1063/1.5012980} {\bibfield  {journal} {\bibinfo  {journal} {J. Chem.
  Phys.}\ }\textbf {\bibinfo {volume} {148}},\ \bibinfo {pages} {241720}
  (\bibinfo {year} {2018})}\BibitemShut {NoStop}%
\bibitem [{\citenamefont {Artrith}(2019)}]{jpe1-2019-ML-for-Interfaces}%
  \BibitemOpen
  \bibfield  {author} {\bibinfo {author} {\bibfnamefont {N.}~\bibnamefont
  {Artrith}},\ }\href {\doibase 10.1088/2515-7655/ab2060} {\bibfield  {journal}
  {\bibinfo  {journal} {J. Phys. Energy}\ }\textbf {\bibinfo {volume} {1}},\
  \bibinfo {pages} {032002} (\bibinfo {year} {2019})}\BibitemShut {NoStop}%
\bibitem [{\citenamefont {Eckhoff}\ and\ \citenamefont
  {Behler}(2019)}]{joctac15-2019-3793}%
  \BibitemOpen
  \bibfield  {author} {\bibinfo {author} {\bibfnamefont {M.}~\bibnamefont
  {Eckhoff}}\ and\ \bibinfo {author} {\bibfnamefont {J.}~\bibnamefont
  {Behler}},\ }\href {\doibase 10.1021/acs.jctc.8b01288} {\bibfield  {journal}
  {\bibinfo  {journal} {J. Chem. Theory Comput.}\ }\textbf {\bibinfo {volume}
  {15}},\ \bibinfo {pages} {3793} (\bibinfo {year} {2019})}\BibitemShut
  {NoStop}%
\bibitem [{\citenamefont {Singraber}\ \emph {et~al.}(2019)\citenamefont
  {Singraber}, \citenamefont {Morawietz}, \citenamefont {Behler},\ and\
  \citenamefont {Dellago}}]{jctc15-2019-3075}%
  \BibitemOpen
  \bibfield  {author} {\bibinfo {author} {\bibfnamefont {A.}~\bibnamefont
  {Singraber}}, \bibinfo {author} {\bibfnamefont {T.}~\bibnamefont
  {Morawietz}}, \bibinfo {author} {\bibfnamefont {J.}~\bibnamefont {Behler}}, \
  and\ \bibinfo {author} {\bibfnamefont {C.}~\bibnamefont {Dellago}},\ }\href
  {\doibase 10.1021/acs.jctc.8b01092} {\bibfield  {journal} {\bibinfo
  {journal} {J. Chem. Theory Comput.}\ }\textbf {\bibinfo {volume} {15}},\
  \bibinfo {pages} {3075} (\bibinfo {year} {2019})}\BibitemShut {NoStop}%
\bibitem [{\citenamefont {Artrith}\ and\ \citenamefont
  {Urban}(2016)}]{cms114-2016-135}%
  \BibitemOpen
  \bibfield  {author} {\bibinfo {author} {\bibfnamefont {N.}~\bibnamefont
  {Artrith}}\ and\ \bibinfo {author} {\bibfnamefont {A.}~\bibnamefont
  {Urban}},\ }\href {\doibase 10.1016/j.commatsci.2015.11.047} {\bibfield
  {journal} {\bibinfo  {journal} {Comput. Mater. Sci.}\ }\textbf {\bibinfo
  {volume} {114}},\ \bibinfo {pages} {135} (\bibinfo {year}
  {2016})}\BibitemShut {NoStop}%
\bibitem [{\citenamefont {Daw}\ and\ \citenamefont
  {Baskes}(1984)}]{prB29-1984-6443}%
  \BibitemOpen
  \bibfield  {author} {\bibinfo {author} {\bibfnamefont {M.~S.}\ \bibnamefont
  {Daw}}\ and\ \bibinfo {author} {\bibfnamefont {M.~I.}\ \bibnamefont
  {Baskes}},\ }\href {\doibase 10.1103/PhysRevB.29.6443} {\bibfield  {journal}
  {\bibinfo  {journal} {Phys. Rev. B}\ }\textbf {\bibinfo {volume} {29}},\
  \bibinfo {pages} {6443} (\bibinfo {year} {1984})}\BibitemShut {NoStop}%
\bibitem [{\citenamefont {Behler}(2011)}]{jcp134-2011-074106}%
  \BibitemOpen
  \bibfield  {author} {\bibinfo {author} {\bibfnamefont {J.}~\bibnamefont
  {Behler}},\ }\href {\doibase 10.1063/1.3553717} {\bibfield  {journal}
  {\bibinfo  {journal} {J. Chem. Phys.}\ }\textbf {\bibinfo {volume} {134}},\
  \bibinfo {pages} {074106} (\bibinfo {year} {2011})}\BibitemShut {NoStop}%
\bibitem [{\citenamefont {Huang}\ \emph {et~al.}(2019)\citenamefont {Huang},
  \citenamefont {Kang}, \citenamefont {Goddard},\ and\ \citenamefont
  {Wang}}]{prb99-2019-064103}%
  \BibitemOpen
  \bibfield  {author} {\bibinfo {author} {\bibfnamefont {Y.}~\bibnamefont
  {Huang}}, \bibinfo {author} {\bibfnamefont {J.}~\bibnamefont {Kang}},
  \bibinfo {author} {\bibfnamefont {W.~A.}\ \bibnamefont {Goddard}}, \ and\
  \bibinfo {author} {\bibfnamefont {L.-W.}\ \bibnamefont {Wang}},\ }\href
  {\doibase 10.1103/physrevb.99.064103} {\bibfield  {journal} {\bibinfo
  {journal} {Phys. Rev. B}\ }\textbf {\bibinfo {volume} {99}},\ \bibinfo
  {pages} {064103} (\bibinfo {year} {2019})}\BibitemShut {NoStop}%
\bibitem [{\citenamefont {LeCun}\ \emph {et~al.}(1998)\citenamefont {LeCun},
  \citenamefont {Bottou}, \citenamefont {Orr},\ and\ \citenamefont
  {Müller}}]{OrrMueller1998_BackProp}%
  \BibitemOpen
  \bibfield  {author} {\bibinfo {author} {\bibfnamefont {Y.}~\bibnamefont
  {LeCun}}, \bibinfo {author} {\bibfnamefont {L.}~\bibnamefont {Bottou}},
  \bibinfo {author} {\bibfnamefont {G.~B.}\ \bibnamefont {Orr}}, \ and\
  \bibinfo {author} {\bibfnamefont {K.-R.}\ \bibnamefont {Müller}},\ }in\
  \href {\doibase 10.1007/3-540-49430-8_2} {\emph {\bibinfo {booktitle} {Neural
  Networks: Tricks of the Trade}}},\ \bibinfo {series} {Lecture Notes in
  Computer Science}, Vol.\ \bibinfo {volume} {1524},\ \bibinfo {editor} {edited
  by\ \bibinfo {editor} {\bibfnamefont {G.~B.}\ \bibnamefont {Orr}}\ and\
  \bibinfo {editor} {\bibfnamefont {K.-R.}\ \bibnamefont {Müller}}}\ (\bibinfo
   {publisher} {Springer Berlin Heidelberg},\ \bibinfo {year} {1998})\ pp.\
  \bibinfo {pages} {9--50}\BibitemShut {NoStop}%
\bibitem [{\citenamefont {Lorenz}(2001)}]{Lorenz2001}%
  \BibitemOpen
  \bibfield  {author} {\bibinfo {author} {\bibfnamefont {S.}~\bibnamefont
  {Lorenz}},\ }\emph {\bibinfo {title} {Reactions on Surfaces with Neural
  Networks}},\ \href
  {http://opus4.kobv.de/opus4-tuberlin/frontdoor/index/index/docId/202}
  {\bibinfo {type} {Phd thesis}},\ \bibinfo  {school} {Technischen
  Universit\"at Berlin, Fakult\"at II -- Mathematik und Naturwissenschaften}
  (\bibinfo {year} {2001})\BibitemShut {NoStop}%
\bibitem [{\citenamefont {Vlcek}, \citenamefont {Sun},\ and\ \citenamefont
  {Kent}(2017)}]{jcp147-2017-161713}%
  \BibitemOpen
  \bibfield  {author} {\bibinfo {author} {\bibfnamefont {L.}~\bibnamefont
  {Vlcek}}, \bibinfo {author} {\bibfnamefont {W.}~\bibnamefont {Sun}}, \ and\
  \bibinfo {author} {\bibfnamefont {P.~R.~C.}\ \bibnamefont {Kent}},\ }\href
  {\doibase 10.1063/1.4986079} {\bibfield  {journal} {\bibinfo  {journal} {J.
  Chem. Phys.}\ }\textbf {\bibinfo {volume} {147}},\ \bibinfo {pages} {161713}
  (\bibinfo {year} {2017})}\BibitemShut {NoStop}%
\bibitem [{\citenamefont {Lennard-Jones}(1929)}]{tfs25-1929-668}%
  \BibitemOpen
  \bibfield  {author} {\bibinfo {author} {\bibfnamefont {J.~E.}\ \bibnamefont
  {Lennard-Jones}},\ }\href {\doibase 10.1039/TF9292500668} {\bibfield
  {journal} {\bibinfo  {journal} {Trans. Faraday Soc.}\ }\textbf {\bibinfo
  {volume} {25}},\ \bibinfo {pages} {668} (\bibinfo {year} {1929})}\BibitemShut
  {NoStop}%
\bibitem [{\citenamefont {Grimme}, \citenamefont {Bannwarth},\ and\
  \citenamefont {Shushkov}(2017)}]{jctc13-2017-1989}%
  \BibitemOpen
  \bibfield  {author} {\bibinfo {author} {\bibfnamefont {S.}~\bibnamefont
  {Grimme}}, \bibinfo {author} {\bibfnamefont {C.}~\bibnamefont {Bannwarth}}, \
  and\ \bibinfo {author} {\bibfnamefont {P.}~\bibnamefont {Shushkov}},\ }\href
  {\doibase 10.1021/acs.jctc.7b00118} {\bibfield  {journal} {\bibinfo
  {journal} {J. Chem. Theory Comput.}\ }\textbf {\bibinfo {volume} {13}},\
  \bibinfo {pages} {1989} (\bibinfo {year} {2017})}\BibitemShut {NoStop}%
\bibitem [{\citenamefont {Marsalek}\ and\ \citenamefont
  {Markland}(2017)}]{jpcl8-2017-1545}%
  \BibitemOpen
  \bibfield  {author} {\bibinfo {author} {\bibfnamefont {O.}~\bibnamefont
  {Marsalek}}\ and\ \bibinfo {author} {\bibfnamefont {T.~E.}\ \bibnamefont
  {Markland}},\ }\href {\doibase 10.1021/acs.jpclett.7b00391} {\bibfield
  {journal} {\bibinfo  {journal} {J. Phys. Chem. Lett.}\ }\textbf {\bibinfo
  {volume} {8}},\ \bibinfo {pages} {1545} (\bibinfo {year} {2017})}\BibitemShut
  {NoStop}%
\bibitem [{\citenamefont {Morawietz}\ \emph {et~al.}(2018)\citenamefont
  {Morawietz}, \citenamefont {Marsalek}, \citenamefont {Pattenaude},
  \citenamefont {Streacker}, \citenamefont {Ben-Amotz},\ and\ \citenamefont
  {Markland}}]{jpcl9-2018-851}%
  \BibitemOpen
  \bibfield  {author} {\bibinfo {author} {\bibfnamefont {T.}~\bibnamefont
  {Morawietz}}, \bibinfo {author} {\bibfnamefont {O.}~\bibnamefont {Marsalek}},
  \bibinfo {author} {\bibfnamefont {S.~R.}\ \bibnamefont {Pattenaude}},
  \bibinfo {author} {\bibfnamefont {L.~M.}\ \bibnamefont {Streacker}}, \bibinfo
  {author} {\bibfnamefont {D.}~\bibnamefont {Ben-Amotz}}, \ and\ \bibinfo
  {author} {\bibfnamefont {T.~E.}\ \bibnamefont {Markland}},\ }\href {\doibase
  10.1021/acs.jpclett.8b00133} {\bibfield  {journal} {\bibinfo  {journal} {J.
  Phys. Chem. Lett.}\ }\textbf {\bibinfo {volume} {9}},\ \bibinfo {pages} {851}
  (\bibinfo {year} {2018})}\BibitemShut {NoStop}%
\bibitem [{\citenamefont {Lee}\ \emph {et~al.}(2015)\citenamefont {Lee},
  \citenamefont {Seo}, \citenamefont {Balasubramanian}, \citenamefont {Twu},
  \citenamefont {Li},\ and\ \citenamefont {Ceder}}]{ees8-2015-3255}%
  \BibitemOpen
  \bibfield  {author} {\bibinfo {author} {\bibfnamefont {J.}~\bibnamefont
  {Lee}}, \bibinfo {author} {\bibfnamefont {D.-H.}\ \bibnamefont {Seo}},
  \bibinfo {author} {\bibfnamefont {M.}~\bibnamefont {Balasubramanian}},
  \bibinfo {author} {\bibfnamefont {N.}~\bibnamefont {Twu}}, \bibinfo {author}
  {\bibfnamefont {X.}~\bibnamefont {Li}}, \ and\ \bibinfo {author}
  {\bibfnamefont {G.}~\bibnamefont {Ceder}},\ }\href {\doibase
  10.1039/c5ee02329g} {\bibfield  {journal} {\bibinfo  {journal} {Energy
  Environ. Sci.}\ }\textbf {\bibinfo {volume} {8}},\ \bibinfo {pages}
  {3255–3265} (\bibinfo {year} {2015})}\BibitemShut {NoStop}%
\bibitem [{\citenamefont {Togo}\ and\ \citenamefont
  {Tanaka}(2015)}]{sm108-2015-1}%
  \BibitemOpen
  \bibfield  {author} {\bibinfo {author} {\bibfnamefont {A.}~\bibnamefont
  {Togo}}\ and\ \bibinfo {author} {\bibfnamefont {I.}~\bibnamefont {Tanaka}},\
  }\href {\doibase 10.1016/j.scriptamat.2015.07.021} {\bibfield  {journal}
  {\bibinfo  {journal} {Scr. Mater.}\ }\textbf {\bibinfo {volume} {108}},\
  \bibinfo {pages} {1} (\bibinfo {year} {2015})}\BibitemShut {NoStop}%
\bibitem [{\citenamefont {Urban}\ \emph {et~al.}(2017)\citenamefont {Urban},
  \citenamefont {Abdellahi}, \citenamefont {Dacek}, \citenamefont {Artrith},\
  and\ \citenamefont {Ceder}}]{prl119-2017-176402}%
  \BibitemOpen
  \bibfield  {author} {\bibinfo {author} {\bibfnamefont {A.}~\bibnamefont
  {Urban}}, \bibinfo {author} {\bibfnamefont {A.}~\bibnamefont {Abdellahi}},
  \bibinfo {author} {\bibfnamefont {S.}~\bibnamefont {Dacek}}, \bibinfo
  {author} {\bibfnamefont {N.}~\bibnamefont {Artrith}}, \ and\ \bibinfo
  {author} {\bibfnamefont {G.}~\bibnamefont {Ceder}},\ }\href {\doibase
  10.1103/PhysRevLett.119.176402} {\bibfield  {journal} {\bibinfo  {journal}
  {Phys. Rev. Lett.}\ }\textbf {\bibinfo {volume} {119}},\ \bibinfo {pages}
  {176402} (\bibinfo {year} {2017})}\BibitemShut {NoStop}%
\bibitem [{\citenamefont {Todorov}\ \emph {et~al.}(2006)\citenamefont
  {Todorov}, \citenamefont {Smith}, \citenamefont {Trachenko},\ and\
  \citenamefont {Dove}}]{jmc16-2006-1911}%
  \BibitemOpen
  \bibfield  {author} {\bibinfo {author} {\bibfnamefont {I.~T.}\ \bibnamefont
  {Todorov}}, \bibinfo {author} {\bibfnamefont {W.}~\bibnamefont {Smith}},
  \bibinfo {author} {\bibfnamefont {K.}~\bibnamefont {Trachenko}}, \ and\
  \bibinfo {author} {\bibfnamefont {M.~T.}\ \bibnamefont {Dove}},\ }\href
  {\doibase 10.1039/B517931A} {\bibfield  {journal} {\bibinfo  {journal} {J.
  Mater. Chem.}\ }\textbf {\bibinfo {volume} {16}},\ \bibinfo {pages} {1911}
  (\bibinfo {year} {2006})}\BibitemShut {NoStop}%
\bibitem [{\citenamefont {Sherwood}\ \emph {et~al.}(2003)\citenamefont
  {Sherwood}, \citenamefont {de~Vries}, \citenamefont {Guest}, \citenamefont
  {Schreckenbach}, \citenamefont {Catlow}, \citenamefont {French},
  \citenamefont {Sokol}, \citenamefont {Bromley}, \citenamefont {Thiel},
  \citenamefont {Turner}, \citenamefont {Billeter}, \citenamefont {Terstegen},
  \citenamefont {Thiel}, \citenamefont {Kendrick}, \citenamefont {Rogers},
  \citenamefont {Casci}, \citenamefont {Watson}, \citenamefont {King},
  \citenamefont {Karlsen}, \citenamefont {Sj{\o}voll}, \citenamefont {Fahmi},
  \citenamefont {Sch{\"a}fer},\ and\ \citenamefont {Lennartz}}]{jms632-2003-1}%
  \BibitemOpen
  \bibfield  {author} {\bibinfo {author} {\bibfnamefont {P.}~\bibnamefont
  {Sherwood}}, \bibinfo {author} {\bibfnamefont {A.~H.}\ \bibnamefont
  {de~Vries}}, \bibinfo {author} {\bibfnamefont {M.~F.}\ \bibnamefont {Guest}},
  \bibinfo {author} {\bibfnamefont {G.}~\bibnamefont {Schreckenbach}}, \bibinfo
  {author} {\bibfnamefont {C.~R.~A.}\ \bibnamefont {Catlow}}, \bibinfo {author}
  {\bibfnamefont {S.~A.}\ \bibnamefont {French}}, \bibinfo {author}
  {\bibfnamefont {A.~A.}\ \bibnamefont {Sokol}}, \bibinfo {author}
  {\bibfnamefont {S.~T.}\ \bibnamefont {Bromley}}, \bibinfo {author}
  {\bibfnamefont {W.}~\bibnamefont {Thiel}}, \bibinfo {author} {\bibfnamefont
  {A.~J.}\ \bibnamefont {Turner}}, \bibinfo {author} {\bibfnamefont
  {S.}~\bibnamefont {Billeter}}, \bibinfo {author} {\bibfnamefont
  {F.}~\bibnamefont {Terstegen}}, \bibinfo {author} {\bibfnamefont
  {S.}~\bibnamefont {Thiel}}, \bibinfo {author} {\bibfnamefont
  {J.}~\bibnamefont {Kendrick}}, \bibinfo {author} {\bibfnamefont {S.~C.}\
  \bibnamefont {Rogers}}, \bibinfo {author} {\bibfnamefont {J.}~\bibnamefont
  {Casci}}, \bibinfo {author} {\bibfnamefont {M.}~\bibnamefont {Watson}},
  \bibinfo {author} {\bibfnamefont {F.}~\bibnamefont {King}}, \bibinfo {author}
  {\bibfnamefont {E.}~\bibnamefont {Karlsen}}, \bibinfo {author} {\bibfnamefont
  {M.}~\bibnamefont {Sj{\o}voll}}, \bibinfo {author} {\bibfnamefont
  {A.}~\bibnamefont {Fahmi}}, \bibinfo {author} {\bibfnamefont
  {A.}~\bibnamefont {Sch{\"a}fer}}, \ and\ \bibinfo {author} {\bibfnamefont
  {C.}~\bibnamefont {Lennartz}},\ }\href {\doibase
  10.1016/s0166-1280(03)00285-9} {\bibfield  {journal} {\bibinfo  {journal} {J.
  Mol. Struct. (THEOCHEM)}\ }\textbf {\bibinfo {volume} {632}},\ \bibinfo
  {pages} {1} (\bibinfo {year} {2003})}\BibitemShut {NoStop}%
\bibitem [{\citenamefont {Metz}\ \emph {et~al.}(2014)\citenamefont {Metz},
  \citenamefont {K{\"a}stner}, \citenamefont {Sokol}, \citenamefont {Keal},\
  and\ \citenamefont {Sherwood}}]{wcms4-2014-101}%
  \BibitemOpen
  \bibfield  {author} {\bibinfo {author} {\bibfnamefont {S.}~\bibnamefont
  {Metz}}, \bibinfo {author} {\bibfnamefont {J.}~\bibnamefont {K{\"a}stner}},
  \bibinfo {author} {\bibfnamefont {A.~A.}\ \bibnamefont {Sokol}}, \bibinfo
  {author} {\bibfnamefont {T.~W.}\ \bibnamefont {Keal}}, \ and\ \bibinfo
  {author} {\bibfnamefont {P.}~\bibnamefont {Sherwood}},\ }\href {\doibase
  10.1002/wcms.1163} {\bibfield  {journal} {\bibinfo  {journal} {WIREs Comput.
  Mol. Sci.}\ }\textbf {\bibinfo {volume} {4}},\ \bibinfo {pages} {101}
  (\bibinfo {year} {2014})}\BibitemShut {NoStop}%
\bibitem [{\citenamefont {Bannwarth}, \citenamefont {Ehlert},\ and\
  \citenamefont {Grimme}(2019)}]{jctc15-2019-1652}%
  \BibitemOpen
  \bibfield  {author} {\bibinfo {author} {\bibfnamefont {C.}~\bibnamefont
  {Bannwarth}}, \bibinfo {author} {\bibfnamefont {S.}~\bibnamefont {Ehlert}}, \
  and\ \bibinfo {author} {\bibfnamefont {S.}~\bibnamefont {Grimme}},\ }\href
  {\doibase 10.1021/acs.jctc.8b01176} {\bibfield  {journal} {\bibinfo
  {journal} {J. Chem. Theory Comput.}\ }\textbf {\bibinfo {volume} {15}},\
  \bibinfo {pages} {1652} (\bibinfo {year} {2019})}\BibitemShut {NoStop}%
\bibitem [{\citenamefont {Nos{\'{e}}}(1984)}]{tjocp81-1984-511}%
  \BibitemOpen
  \bibfield  {author} {\bibinfo {author} {\bibfnamefont {S.}~\bibnamefont
  {Nos{\'{e}}}},\ }\href {\doibase 10.1063/1.447334} {\bibfield  {journal}
  {\bibinfo  {journal} {J. Chem. Phys.}\ }\textbf {\bibinfo {volume} {81}},\
  \bibinfo {pages} {511} (\bibinfo {year} {1984})}\BibitemShut {NoStop}%
\bibitem [{\citenamefont {Hoover}(1985)}]{pra31-1985-1695}%
  \BibitemOpen
  \bibfield  {author} {\bibinfo {author} {\bibfnamefont {W.~G.}\ \bibnamefont
  {Hoover}},\ }\href {\doibase 10.1103/physreva.31.1695} {\bibfield  {journal}
  {\bibinfo  {journal} {Phys. Rev. A}\ }\textbf {\bibinfo {volume} {31}},\
  \bibinfo {pages} {1695} (\bibinfo {year} {1985})}\BibitemShut {NoStop}%
\bibitem [{\citenamefont {Becke}(1988)}]{pra38-1988-3098}%
  \BibitemOpen
  \bibfield  {author} {\bibinfo {author} {\bibfnamefont {A.~D.}\ \bibnamefont
  {Becke}},\ }\href {\doibase 10.1103/PhysRevA.38.3098} {\bibfield  {journal}
  {\bibinfo  {journal} {Phys. Rev. A}\ }\textbf {\bibinfo {volume} {38}},\
  \bibinfo {pages} {3098} (\bibinfo {year} {1988})}\BibitemShut {NoStop}%
\bibitem [{\citenamefont {Lee}, \citenamefont {Yang},\ and\ \citenamefont
  {Parr}(1988)}]{prb37-1988-785}%
  \BibitemOpen
  \bibfield  {author} {\bibinfo {author} {\bibfnamefont {C.}~\bibnamefont
  {Lee}}, \bibinfo {author} {\bibfnamefont {W.}~\bibnamefont {Yang}}, \ and\
  \bibinfo {author} {\bibfnamefont {R.~G.}\ \bibnamefont {Parr}},\ }\href
  {\doibase 10.1103/PhysRevB.37.785} {\bibfield  {journal} {\bibinfo  {journal}
  {Phys. Rev. B}\ }\textbf {\bibinfo {volume} {37}},\ \bibinfo {pages} {785}
  (\bibinfo {year} {1988})}\BibitemShut {NoStop}%
\bibitem [{\citenamefont {Johnson}, \citenamefont {Gill},\ and\ \citenamefont
  {Pople}(1993)}]{jcp98-1993-5612}%
  \BibitemOpen
  \bibfield  {author} {\bibinfo {author} {\bibfnamefont {B.~G.}\ \bibnamefont
  {Johnson}}, \bibinfo {author} {\bibfnamefont {P.~M.~W.}\ \bibnamefont
  {Gill}}, \ and\ \bibinfo {author} {\bibfnamefont {J.~A.}\ \bibnamefont
  {Pople}},\ }\href {\doibase 10.1063/1.464906} {\bibfield  {journal} {\bibinfo
   {journal} {J. Chem. Phys.}\ }\textbf {\bibinfo {volume} {98}},\ \bibinfo
  {pages} {5612} (\bibinfo {year} {1993})}\BibitemShut {NoStop}%
\bibitem [{\citenamefont {Russo}, \citenamefont {Martin},\ and\ \citenamefont
  {Hay}(1994)}]{jcp101-1994-7729}%
  \BibitemOpen
  \bibfield  {author} {\bibinfo {author} {\bibfnamefont {T.~V.}\ \bibnamefont
  {Russo}}, \bibinfo {author} {\bibfnamefont {R.~L.}\ \bibnamefont {Martin}}, \
  and\ \bibinfo {author} {\bibfnamefont {P.~J.}\ \bibnamefont {Hay}},\ }\href
  {\doibase 10.1063/1.468265} {\bibfield  {journal} {\bibinfo  {journal} {J.
  Chem. Phys.}\ }\textbf {\bibinfo {volume} {101}},\ \bibinfo {pages} {7729}
  (\bibinfo {year} {1994})}\BibitemShut {NoStop}%
\bibitem [{\citenamefont {Grimme}\ \emph {et~al.}(2010)\citenamefont {Grimme},
  \citenamefont {Antony}, \citenamefont {Ehrlich},\ and\ \citenamefont
  {Krieg}}]{jcp132-2010-154104}%
  \BibitemOpen
  \bibfield  {author} {\bibinfo {author} {\bibfnamefont {S.}~\bibnamefont
  {Grimme}}, \bibinfo {author} {\bibfnamefont {J.}~\bibnamefont {Antony}},
  \bibinfo {author} {\bibfnamefont {S.}~\bibnamefont {Ehrlich}}, \ and\
  \bibinfo {author} {\bibfnamefont {H.}~\bibnamefont {Krieg}},\ }\href
  {\doibase 10.1063/1.3382344} {\bibfield  {journal} {\bibinfo  {journal} {J.
  Chem. Phys.}\ }\textbf {\bibinfo {volume} {132}},\ \bibinfo {pages} {154104}
  (\bibinfo {year} {2010})}\BibitemShut {NoStop}%
\bibitem [{\citenamefont {Smith}\ \emph {et~al.}(2016)\citenamefont {Smith},
  \citenamefont {Burns}, \citenamefont {Patkowski},\ and\ \citenamefont
  {Sherrill}}]{jpcl7-2016-2197}%
  \BibitemOpen
  \bibfield  {author} {\bibinfo {author} {\bibfnamefont {D.~G.~A.}\
  \bibnamefont {Smith}}, \bibinfo {author} {\bibfnamefont {L.~A.}\ \bibnamefont
  {Burns}}, \bibinfo {author} {\bibfnamefont {K.}~\bibnamefont {Patkowski}}, \
  and\ \bibinfo {author} {\bibfnamefont {C.~D.}\ \bibnamefont {Sherrill}},\
  }\href {\doibase 10.1021/acs.jpclett.6b00780} {\bibfield  {journal} {\bibinfo
   {journal} {The Journal of Physical Chemistry Letters}\ }\textbf {\bibinfo
  {volume} {7}},\ \bibinfo {pages} {2197} (\bibinfo {year} {2016})}\BibitemShut
  {NoStop}%
\bibitem [{\citenamefont {Furche}\ \emph {et~al.}(2014)\citenamefont {Furche},
  \citenamefont {Ahlrichs}, \citenamefont {Hättig}, \citenamefont {Klopper},
  \citenamefont {Sierka},\ and\ \citenamefont {Weigend}}]{ircms4-2014-91}%
  \BibitemOpen
  \bibfield  {author} {\bibinfo {author} {\bibfnamefont {F.}~\bibnamefont
  {Furche}}, \bibinfo {author} {\bibfnamefont {R.}~\bibnamefont {Ahlrichs}},
  \bibinfo {author} {\bibfnamefont {C.}~\bibnamefont {Hättig}}, \bibinfo
  {author} {\bibfnamefont {W.}~\bibnamefont {Klopper}}, \bibinfo {author}
  {\bibfnamefont {M.}~\bibnamefont {Sierka}}, \ and\ \bibinfo {author}
  {\bibfnamefont {F.}~\bibnamefont {Weigend}},\ }\href {\doibase
  10.1002/wcms.1162} {\bibfield  {journal} {\bibinfo  {journal} {WIREs Comput
  Mol Sci}\ }\textbf {\bibinfo {volume} {4}},\ \bibinfo {pages} {91} (\bibinfo
  {year} {2014})}\BibitemShut {NoStop}%
\bibitem [{\citenamefont {Kresse}\ and\ \citenamefont
  {Furthm\"uller}(1996{\natexlab{a}})}]{prB54-1996-11169}%
  \BibitemOpen
  \bibfield  {author} {\bibinfo {author} {\bibfnamefont {G.}~\bibnamefont
  {Kresse}}\ and\ \bibinfo {author} {\bibfnamefont {J.}~\bibnamefont
  {Furthm\"uller}},\ }\href {\doibase 10.1103/PhysRevB.54.11169} {\bibfield
  {journal} {\bibinfo  {journal} {Phys. Rev. B}\ }\textbf {\bibinfo {volume}
  {54}},\ \bibinfo {pages} {11169} (\bibinfo {year}
  {1996}{\natexlab{a}})}\BibitemShut {NoStop}%
\bibitem [{\citenamefont {Kresse}\ and\ \citenamefont
  {Furthm\"uller}(1996{\natexlab{b}})}]{cms6-1996-15}%
  \BibitemOpen
  \bibfield  {author} {\bibinfo {author} {\bibfnamefont {G.}~\bibnamefont
  {Kresse}}\ and\ \bibinfo {author} {\bibfnamefont {J.}~\bibnamefont
  {Furthm\"uller}},\ }\href {\doibase 10.1016/0927-0256(96)00008-0} {\bibfield
  {journal} {\bibinfo  {journal} {Comput. Mater. Sci.}\ }\textbf {\bibinfo
  {volume} {6}},\ \bibinfo {pages} {15} (\bibinfo {year}
  {1996}{\natexlab{b}})}\BibitemShut {NoStop}%
\bibitem [{\citenamefont {Bl\"ochl}(1994)}]{prB50-1994-17953}%
  \BibitemOpen
  \bibfield  {author} {\bibinfo {author} {\bibfnamefont {P.~E.}\ \bibnamefont
  {Bl\"ochl}},\ }\href {\doibase 10.1103/PhysRevB.50.17953} {\bibfield
  {journal} {\bibinfo  {journal} {Phys. Rev. B}\ }\textbf {\bibinfo {volume}
  {50}},\ \bibinfo {pages} {17953} (\bibinfo {year} {1994})}\BibitemShut
  {NoStop}%
\bibitem [{\citenamefont {Zhang}\ and\ \citenamefont
  {Yang}(1998)}]{prl80-1998-890}%
  \BibitemOpen
  \bibfield  {author} {\bibinfo {author} {\bibfnamefont {Y.}~\bibnamefont
  {Zhang}}\ and\ \bibinfo {author} {\bibfnamefont {W.}~\bibnamefont {Yang}},\
  }\href {\doibase 10.1103/physrevlett.80.890} {\bibfield  {journal} {\bibinfo
  {journal} {Phys. Rev. Lett.}\ }\textbf {\bibinfo {volume} {80}},\ \bibinfo
  {pages} {890} (\bibinfo {year} {1998})}\BibitemShut {NoStop}%
\bibitem [{\citenamefont {Sun}, \citenamefont {Ruzsinszky},\ and\ \citenamefont
  {Perdew}(2015)}]{prl115-2015-036402}%
  \BibitemOpen
  \bibfield  {author} {\bibinfo {author} {\bibfnamefont {J.}~\bibnamefont
  {Sun}}, \bibinfo {author} {\bibfnamefont {A.}~\bibnamefont {Ruzsinszky}}, \
  and\ \bibinfo {author} {\bibfnamefont {J.}~\bibnamefont {Perdew}},\ }\href
  {\doibase 10.1103/physrevlett.115.036402} {\bibfield  {journal} {\bibinfo
  {journal} {Phys. Rev. Lett.}\ }\textbf {\bibinfo {volume} {115}},\ \bibinfo
  {pages} {036402} (\bibinfo {year} {2015})}\BibitemShut {NoStop}%
\bibitem [{\citenamefont {Byrd}\ \emph {et~al.}(1995)\citenamefont {Byrd},
  \citenamefont {Lu}, \citenamefont {Nocedal},\ and\ \citenamefont
  {Zhu}}]{siam16-1995-1190}%
  \BibitemOpen
  \bibfield  {author} {\bibinfo {author} {\bibfnamefont {R.}~\bibnamefont
  {Byrd}}, \bibinfo {author} {\bibfnamefont {P.}~\bibnamefont {Lu}}, \bibinfo
  {author} {\bibfnamefont {J.}~\bibnamefont {Nocedal}}, \ and\ \bibinfo
  {author} {\bibfnamefont {C.}~\bibnamefont {Zhu}},\ }\href {\doibase
  10.1137/0916069} {\bibfield  {journal} {\bibinfo  {journal} {SIAM Journal on
  Scientific Computing}\ }\textbf {\bibinfo {volume} {16}},\ \bibinfo {pages}
  {1190} (\bibinfo {year} {1995})}\BibitemShut {NoStop}%
\bibitem [{\citenamefont {Zhu}\ \emph {et~al.}(1997)\citenamefont {Zhu},
  \citenamefont {Byrd}, \citenamefont {Lu},\ and\ \citenamefont
  {Nocedal}}]{toms23-1997-550}%
  \BibitemOpen
  \bibfield  {author} {\bibinfo {author} {\bibfnamefont {C.}~\bibnamefont
  {Zhu}}, \bibinfo {author} {\bibfnamefont {R.~H.}\ \bibnamefont {Byrd}},
  \bibinfo {author} {\bibfnamefont {P.}~\bibnamefont {Lu}}, \ and\ \bibinfo
  {author} {\bibfnamefont {J.}~\bibnamefont {Nocedal}},\ }\href {\doibase
  10.1145/279232.279236} {\bibfield  {journal} {\bibinfo  {journal} {ACM T.
  Math Software}\ }\textbf {\bibinfo {volume} {23}},\ \bibinfo {pages}
  {550–560} (\bibinfo {year} {1997})}\BibitemShut {NoStop}%
\bibitem [{\citenamefont {Ponder}\ and\ \citenamefont
  {Richards}(1987)}]{jocc8-1987-1016}%
  \BibitemOpen
  \bibfield  {author} {\bibinfo {author} {\bibfnamefont {J.~W.}\ \bibnamefont
  {Ponder}}\ and\ \bibinfo {author} {\bibfnamefont {F.~M.}\ \bibnamefont
  {Richards}},\ }\href {\doibase 10.1002/jcc.540080710} {\bibfield  {journal}
  {\bibinfo  {journal} {J. Comput. Chem.}\ }\textbf {\bibinfo {volume} {8}},\
  \bibinfo {pages} {1016–1024} (\bibinfo {year} {1987})}\BibitemShut
  {NoStop}%
\bibitem [{\citenamefont {Verlet}(1967)}]{pr159-67-98.pdf}%
  \BibitemOpen
  \bibfield  {author} {\bibinfo {author} {\bibfnamefont {L.}~\bibnamefont
  {Verlet}},\ }\href {\doibase 10.1103/PhysRev.159.98} {\bibfield  {journal}
  {\bibinfo  {journal} {Phys. Rev.}\ }\textbf {\bibinfo {volume} {159}},\
  \bibinfo {pages} {98} (\bibinfo {year} {1967})}\BibitemShut {NoStop}%
\bibitem [{\citenamefont {Bussi}, \citenamefont {Donadio},\ and\ \citenamefont
  {Parrinello}(2007)}]{jcp126-2007-14101}%
  \BibitemOpen
  \bibfield  {author} {\bibinfo {author} {\bibfnamefont {G.}~\bibnamefont
  {Bussi}}, \bibinfo {author} {\bibfnamefont {D.}~\bibnamefont {Donadio}}, \
  and\ \bibinfo {author} {\bibfnamefont {M.}~\bibnamefont {Parrinello}},\
  }\href {\doibase 10.1063/1.2408420} {\bibfield  {journal} {\bibinfo
  {journal} {J. Chem. Phys.}\ }\textbf {\bibinfo {volume} {126}},\ \bibinfo
  {pages} {014101} (\bibinfo {year} {2007})}\BibitemShut {NoStop}%
\bibitem [{\citenamefont {Kokalj}(1999)}]{jmgm17-1999-176}%
  \BibitemOpen
  \bibfield  {author} {\bibinfo {author} {\bibfnamefont {A.}~\bibnamefont
  {Kokalj}},\ }\href {\doibase 10.1016/s1093-3263(99)00028-5} {\bibfield
  {journal} {\bibinfo  {journal} {J. Mol. Graphics Modell.}\ }\textbf {\bibinfo
  {volume} {17}},\ \bibinfo {pages} {176–179} (\bibinfo {year}
  {1999})}\BibitemShut {NoStop}%
\end{thebibliography}%

\newpage
\clearpage
\appendix

\onecolumngrid
\renewcommand{\thefigure}{S\arabic{figure}}
\renewcommand{\thetable}{S\arabic{table}}
\setcounter{figure}{0}
\setcounter{table}{0}

\begin{center}
\LARGE
Supplementary Information
\end{center}

\section*{Supplementary Tables}

\begin{table}[H]
\small
\caption{Optimal parameters for applying the Taylor-expansion approach
  with displacement strategies (A) and (B) to the water-cluster system
  with six water molecules.}
  \label{tbl:rates_inc_H }
  \begin{tabular*}{0.98\textwidth}{@{\extracolsep{\fill}}l c@{ }c@{}c@{}c}
    \toprule
    \textbf{Reference data set}
    & \multicolumn{2}{l}{\textbf{Displacement strategy~(A)}}
    & \multicolumn{2}{l}{\textbf{Displacement strategy~(B)}}\\
    & $\delta$ [\AA{}]& multiple $a$ & $\delta_\textup{max}$ [\AA{}] & multiple $a$\\
    \midrule
    train\_0500 & 0.03 & 11  & 0.008 & 22 \\
    train\_1000 & 0.04 & 32  & 0.010 & 32 \\
    train\_2000 & 0.04 & 32  & 0.010 & 32 \\
    \bottomrule
  \end{tabular*}
\end{table}

\section*{Supplementary Figures}

\begin{figure}[H]
\centering
 \includegraphics[width=0.8\textwidth]{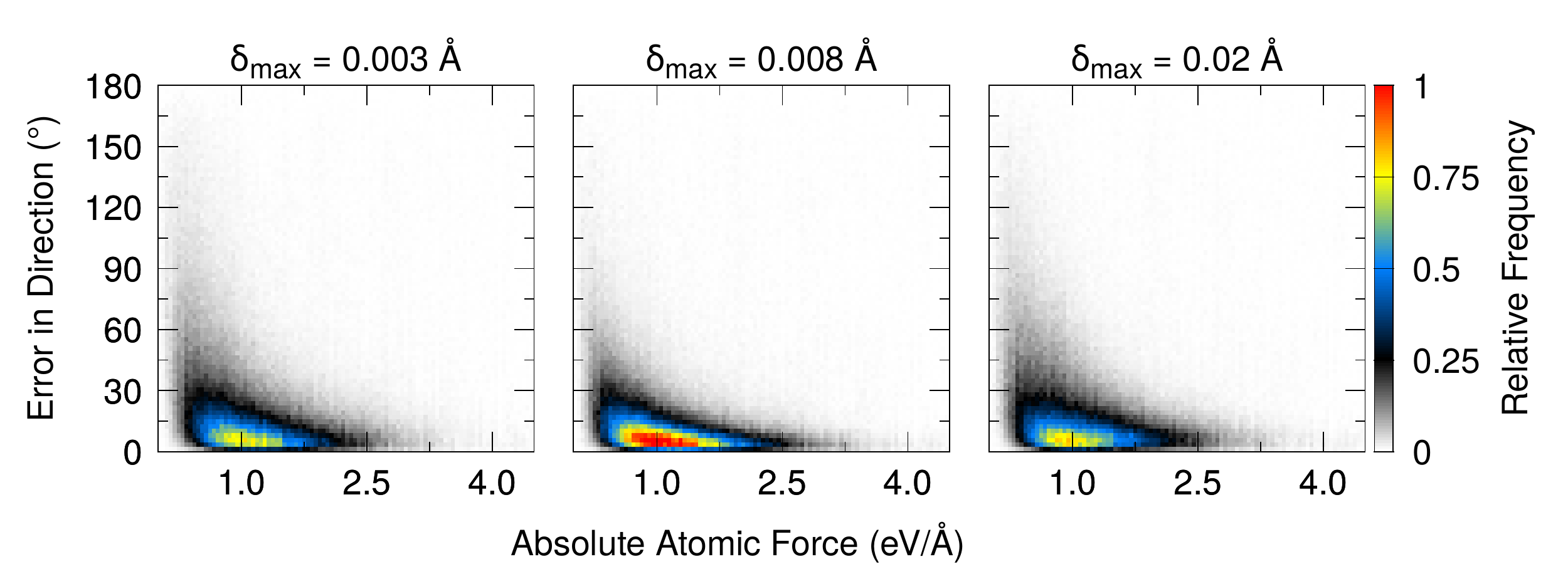}
 \caption{\label{fig:heatmap_delta}%
   Robustness of the improvement of the prediction of the direction of
   the forces with respect to different choices of the maximum
   displacement $\delta_{\textup{max}}$ for the water-cluster system
   with six water molecules.  The error was obtained from 10 ANN
   potentials per displacement that were trained using the train\_0500
   data set employing displacement strategy(ii) for the fraction value
   optimal for the respective delta.}
\end{figure}

\begin{figure}[H]
\centering
 \includegraphics[width=0.45\textwidth]{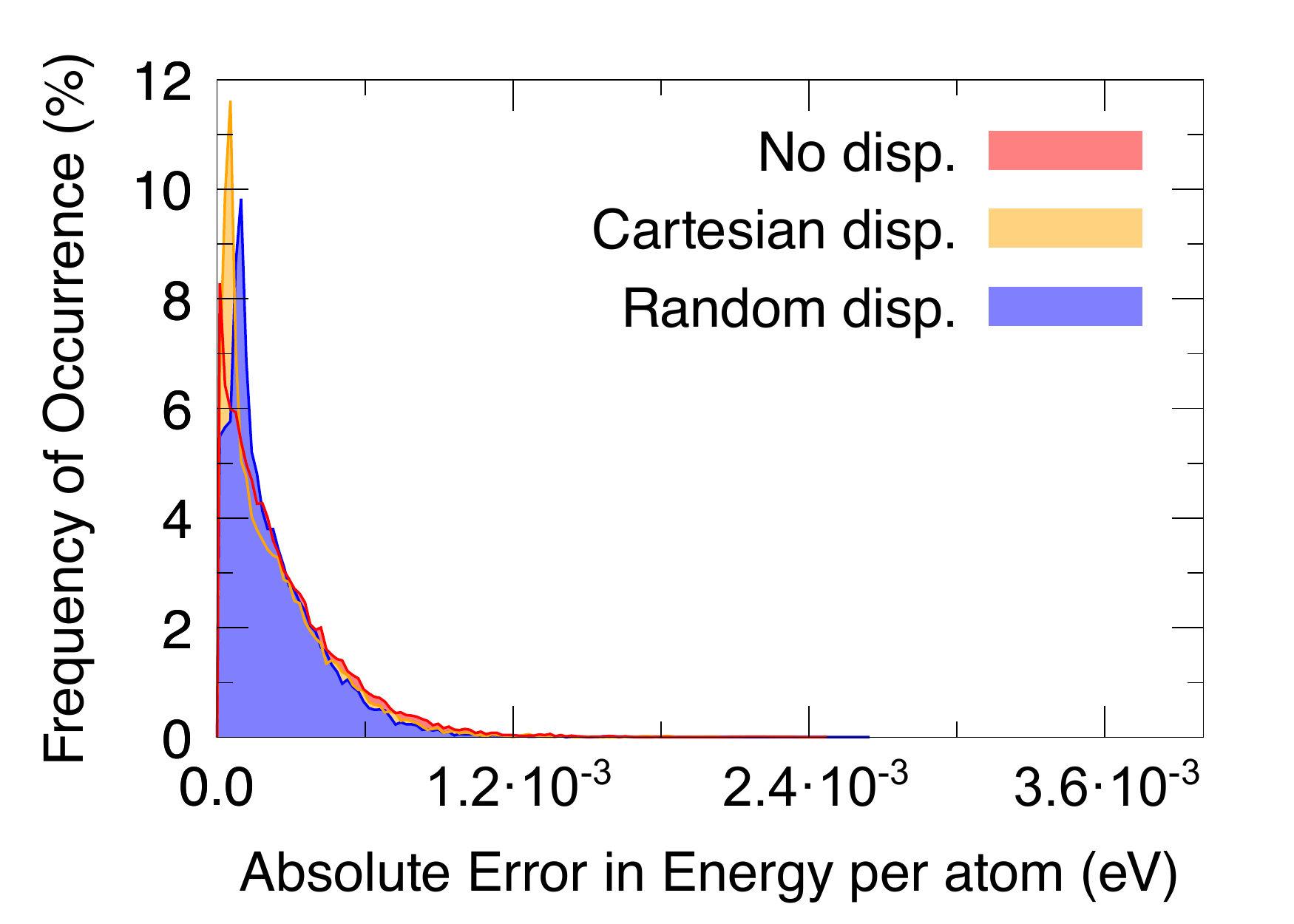}
 \caption{\label{fig:waterc_energy_hist_testset}%
   Histogram of the absolute error in the prediction of the energy per
   atom for the structures in the training data set (train\_0500) for
   the water cluster system.  The respective optimal parameters for the
   displacement and fraction where chosen for both displacement
   strategies.  }
\end{figure}

\begin{figure}[H]
\centering
 \includegraphics[width=0.6\textwidth]{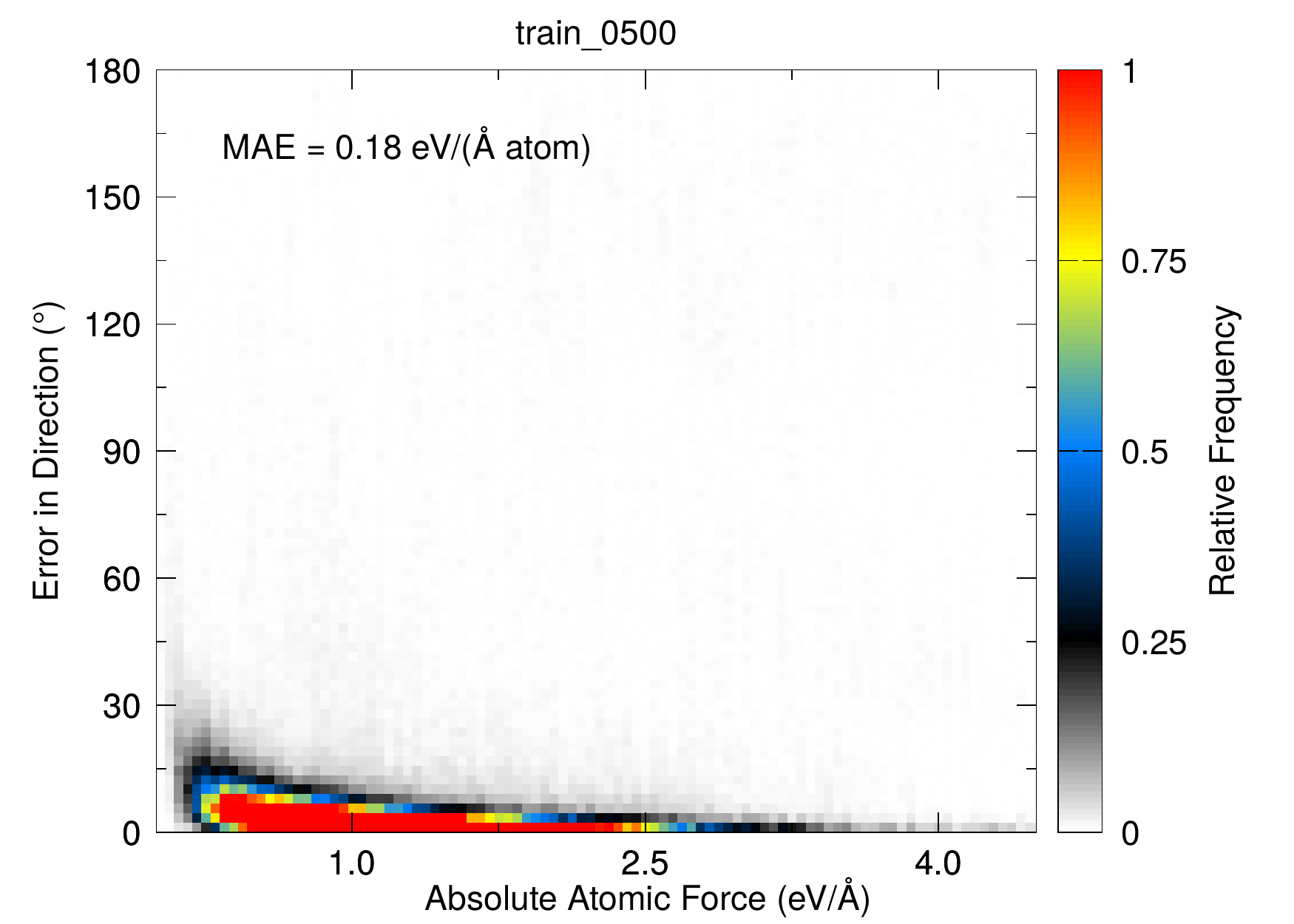}
 \caption{\label{fig:heatmap_delta}%
   Heatmap of the error in the direction of the force vector obtained
   from 10 ANN potentials fitted with the direct force training method
   on the water cluster data set.}
\end{figure}

\end{document}